\documentstyle[aps,prb,preprint,tighten,floats,epsf]{revtex}
\begin{document}
\draft \title{Bloch oscillations of magnetic solitons\\ in anisotropic
spin-1/2 chains} \author{Jordan Kyriakidis\cite{JK.E-mail} and Daniel
Loss\cite{DL.E-mail}} \address{Department of Physics and Astronomy,
University of Basel,\\ Klingelbergstrasse 82, CH--4056 Basel,
Switzerland} \date{\today} \maketitle
\begin{abstract}
We study the quantum dynamics of soliton-like domain walls in
anisotropic spin-1/2 chains in the presence of magnetic fields.  In
the absence of fields, domain walls form a Bloch band of delocalized
quantum states while a static field applied along the easy axis
localizes them into Wannier wave packets and causes them to execute
Bloch oscillations, i.e.\ the domain walls oscillate along the chain
with a finite Bloch frequency and amplitude.  In the presence of the
field, the Bloch band, with a continuum of extended states, breaks up
into the Wannier-Zeeman ladder---a discrete set of equally spaced
energy levels.  We calculate the dynamical structure factor $S^{zz}
(q,\omega)$ in the one-soliton sector at finite frequency, wave
vector, and temperature, and find sharp peaks at frequencies which are
integer multiples of the Bloch frequency.  We further calculate the
uniform magnetic susceptibility and find that it too exhibits peaks at
the Bloch frequency.  We identify several candidate materials where
these Bloch oscillations should be observable, for example, via
neutron scattering measurements. For the particular compound ${\rm
CoCl_2 \!  \cdot \!  2H_2O}$ we estimate the Bloch amplitude to be on
the order of a few lattice constants, and the Bloch frequency on the
order of 100\,GHz for magnetic fields in the Tesla range and at
temperatures of about 18\,Kelvin.
\end{abstract}

\section{Introduction}

Bloch oscillations\cite{bloch28,wannier62,madelung} of a quantum state
in a periodic one-dimensional structure can be characterized as an
{\em oscillatory} response to a {\em constant} force.  The phenomenon
is a remarkable example of the counter-intuitive nature of quantum
mechanics.  Not only does the particle oscillate in response to a
static and homogeneous force, but the amplitude of the oscillation is
{\em inversely} proportional to the magnitude of the force.

The prototypical system in which to observe Bloch oscillations ({\sc
bo}) has historically been a band electron in an external electric
field.\cite{bloch28,wannier62,madelung} In the absence of inelastic
scattering and interband (Zener) transitions, the electron, in
momentum space, continually traverses the Brillouin zone.  Upon
reaching a zone boundary, the electron is reflected to the opposite
boundary and thereby reverses its momentum.  In real space, the motion
is likewise periodic---a constant force thus produces oscillatory
motion.  But the existence of {\sc bo} has been controversial ever
since its theoretical prediction many decades ago
(Ref.~\onlinecite{wannier62} contains an early review).  The
controversy surrounding the theory was largely concerned with the use,
or rather the misuse, of Bloch's theorem in systems where the lattice
periodicity was explicitly broken by a constant external field.
Experiments were not conclusive because, in these early times, bulk
solids were the only systems available. Their bandwidths, however, are
typically too large, the lattice constants too small, and the
inelastic scattering so frequent that the phase coherence of the
particle state, essential for the existence of {\sc bo}, is rapidly
destroyed.

However, with the advent of semiconductor heterostructures---in
particular layered superlattices with large lattice constants, and
thus small bandwidths, which allow many coherent {\sc bo} before phase
coherence is lost---the debate about {\sc bo} seems to be settled.
Recent experiments now provide convincing evidence that {\sc bo} of
electrons can indeed occur.  This confirmation comes from the
detection of the coherent radiation emitted as ensembles of electrons
execute {\sc bo}.\cite{feldmann92} There is also work which directly
measures the physical displacement of electrons in superlattices as
they oscillate.\cite{lyssenko97}

The search for {\sc bo} has not been limited to electrons.  Recently,
the effect was observed with ultra-cold atoms placed in an optical
standing wave.\cite{dahan96} The standing wave served as the periodic
potential, and a force was simulated by accelerating this periodic
potential.  The momentum distribution of the atoms had the time
dependence expected from the theory of {\sc bo}.

Current studies of {\sc bo} are moving beyond the establishment of
their existence.  In atom-optical systems, {\sc bo} are being used to
test other aspects of quantum theory.  For example, nonexponential
decay of unstable systems was very recently observed in the same
systems used to observe {\sc bo} and the Wannier-Stark
ladder.\cite{wilkinson97} In electronic systems, {\sc bo} are being
studied as a means of producing fast emitters of coherent radiation.
The radiation emitted by the oscillating charges can be tuned over a
wide range of frequencies simply by changing the electric field
strength.  Typical wavelengths are in the submillimeter range.

An entirely different class of systems in which such coherence
phenomena can be expected are magnetic systems, which brings us to the
main subject of this work.  It has been pointed out recently that {\sc
bo} should exist in purely magnetic systems---in the quantum dynamics
of domain walls with soliton-like behavior.\cite{braun94} This
proposal for magnetic {\sc bo} was based on a semiclassical
treatment\cite{braun96} of the quantum dynamics of extended domain
walls moving in a periodic potential and containing a large number of
spins $s$, with $s\gg 1$.  In the present work, we shall extend this
investigation to the fully quantum mechanical regime of anisotropic
spin-1/2 chains and demonstrate that {\sc bo} occur in the quantum
dynamics of elementary excitations such as spin solitons.  Such
solitons represent the extreme limit of a magnetic domain wall with a
width of only one lattice constant.  In contrast to the cases
mentioned above (electrons and atoms), magnetic {\sc bo} are an
inherently many-body effect.  The soliton motion is a cooperative
phenomenon resulting from spin-spin interaction.  Nevertheless, we
will see that magnetic {\sc bo} share many properties with their
electronic counterpart.

A remarkable feature of magnetic {\sc bo} is that they give rise to
oscillations of the magnetization at a Bloch frequency which can be
continuously varied by an external magnetic field.  Thus, besides
being of fundamental interest, {\sc bo} of magnetic solitons may also
prove relevant for applications since they provide a natural source of
{\em magnetic} dipole radiation---typically in the microwave regime.

The outline of this work is as follows.  In the next section, we
define and discuss the one-soliton approximation in the presence of a
magnetic field $B$ applied along the easy axis, thereby extending the
zero-field results obtained previously for antiferromagnets ({\sc
afm}s)\cite{villain75} and for ferromagnets ({\sc
fm}s).\cite{braun96a} We show that the external $B$ field localizes
the eigenstates of the Bloch band and discretizes the spectrum into a
set of equally spaced levels which we call the Wannier-Zeeman ladder
({\sc wzl}), in analogy to the Wannier-Stark ladder in electronic
systems.  We mention several spin-models capable of supporting {\sc
bo}, but our main focus is on Ising-like {\sc fm} chains with biaxial
anisotropy---a model which accurately describes compounds such
as\cite{torrance69} ${\rm CoCl_2 \! \cdot \!  2H_2O}$.  We then go on
to discuss the spin correlation functions and calculate the wave
vector and frequency dependent dynamical structure factor $S^{zz}
(q,\omega)$ for two cases: zero and finite magnetic fields.  For
finite magnetic fields, the structure factor consists of peaks at
frequencies corresponding to integer multiples of the Bloch frequency.
These frequencies are typically in the GHz range.  This result means
that, for example, neutron scattering measurements on samples in
thermal equilibrium can be used to observe the {\sc wzl}.  In zero
magnetic field, we obtain a structure factor for {\sc fm}s that is
very similar in form to the one obtained for {\sc
afm}s,\cite{villain75} apart from a wave-vector dependence that
reflects the difference in magnetic ordering between {\sc fm}s and
{\sc afm}s.  After a brief digression on the $B \rightarrow 0$ limit,
we turn to an investigation of specific materials which are promising
candidates in which to observe {\sc bo} of magnetic solitons.  We
focus on one particular Ising-like {\sc fm},\cite{torrance69} ${\rm
CoCl_2 \! \cdot \! 2H_2O}$, which appears to be the best characterized
of the ones we have identified.  Other materials, though less well
characterized, are potentially better candidates.  Finally, we close
with a summary of the main results, along with an outlook on future
directions.

\section{The One-Soliton Approximation}
\label{sec-1.sol.approx}

In the following we concentrate on one-dimensional Ising-like magnets
with nearest-neighbor exchange interactions $J^\alpha$ and in the
presence of a magnetic field ${\bf b}_n = g \mu_B {\bf B}_n$.  The
general anisotropic spin Hamiltonian is given by
\begin{equation}
H = - \sum_{n,\alpha} \bigl( J^\alpha S_n^\alpha S_{n+1}^\alpha +
b_n^\alpha S_n^\alpha \bigr),
\label{eq-h.general}
\end{equation}
where $n$ denotes lattice sites and $\alpha$ denotes Cartesian
coordinates.  The exchange constants $J^\alpha$ are either positive
({\sc fm}s) or negative ({\sc afm}s), with $|J^z| \gg |J^{x,y}|$.
Thus, the (Ising) easy-axis is along the $z$-axis.  In systems with
such a strong easy-axis, domain walls, or solitons (we shall use these
terms interchangeably), are well defined.  At sufficiently low
temperatures, the system is in its ground state: ferromagnetic order
for {\sc fm}s, and N\'eel order for {\sc afm}s.  The excitations
consist of domain walls.  In a pure Ising chain ($J^x = J^y = 0$) with
zero field ($b_n^\alpha = 0$), the spectrum consists of discrete
energy levels, where each level corresponds to states with a fixed
number of domain walls.  If, as is usually the case, there are
additional exchange couplings in the directions transverse to the
Ising axis ($J^x \neq 0$ or $J^y \neq 0$), then the degeneracy is
lifted.  The energy spectrum consists of a series of continuua
separated by gaps.  Each continuum consists of states with a fixed
number of domain walls.  The one-soliton approximation considers only
the lowest band and neglects all transitions to higher bands.
   
The spectrum described in the preceding paragraph has been verified
numerically\cite{mikeska91a} for the anisotropic $x$-$y$ model; for
very large anisotropy (near the Ising limit), isolated bands with
large gaps were observed, with the gap tending to zero as the
isotropic limit was approached.  The one-soliton approximation was
used by Villain\cite{villain75} in his pioneering work on spin-1/2
solitons in Ising-like {\sc afm} chains.  The result of his
calculation---the existence of a dispersive soliton mode (the Villain
mode) below the two particle continuum---was further verified by
theoretical\cite{mikeska} and numerical\cite{ishimura80} work,
inelastic neutron scattering experiments on the Ising-like {\sc afm}s
${\rm CsCoCl_3}$ (Refs.~\onlinecite{yoshizawa81,boucher85}) and ${\rm
CsCoBr_3}$ (Refs.~\onlinecite{nagler82,nagler83}), as well as electron
spin resonance,\cite{adachi81} nuclear magnetic
resonance,\cite{ajiro89} and optical\cite{mogi87} experiments.

Analytic work guided by physical reasoning, exact numerical work on
finite systems, and experimental work on real physical systems all
point to the existence of the Villain mode and thus justify the
one-soliton approximation. What then are the conditions necessary for
the Villain mode to exist?  Two conditions should be met.  First, a
large easy-axis is required; the system must be near the Ising limit.
This ensures not only that domain walls are well-defined excitations,
but also that states with different numbers of domain walls are well
separated in energy.  Second, the domain walls must have dynamics.  It
is the dynamics of the domain walls which induce the band structure.
In a pure Ising chain, for example, a localized domain wall is an
eigenstate of the Hamiltonian and therefore has no dynamics.  The
Villain mode does not exist in pure Ising chains.

Most of the existing work has focused so far on {\sc afm}s, and only
recently has it been pointed out that dispersive soliton modes can
also exist in Ising-like {\sc fm}s\cite{braun96a} and other spin
chains, as discussed next.

\subsection{Primary Model}

In view of the candidate materials to be identified in
Sec.~\ref{sec-exp}, we focus mainly on spin-1/2 {\sc fm}s with a
Hamiltonian given by~(\ref{eq-h.general}) with $J^\alpha >
0$.\cite{braun96a} We consider a static and homogeneous field $b$
along the $z$-axis (the Ising direction) and rewrite the Hamiltonian
in a more suggestive form:
\begin{mathletters} \label{eq-h.primary}
\begin{equation}
H = H^I + H^a + H^\perp,
\end{equation} \begin{equation} \label{eq-h.i}
H^I = -J^z \sum_n S^z_n S^z_{n+1} - b \sum_n S^z_n,
\end{equation} \begin{equation} \label{eq-h.a}
H^a = -{1 \over 4} (J^x - J^y) \sum_n (S_n^+ S_{n+1}^+ + S_n^-
S_{n+1}^- ),
\end{equation} \begin{equation} \label{eq-h.perp}
H^\perp = -{1 \over 4} (J^x + J^y) \sum_n (S_n^+ S_{n+1}^- + S_n^-
S_{n + 1}^+),
\end{equation}
\end{mathletters}
where $S_n^\pm = S_n^x \pm i S_n^y$ are the usual raising and lowering
operators.  In preparation for the one-soliton approximation, we
introduce the one-soliton states $\{ |m,Q \rangle \}$, defined by
\begin{equation} \label{eq-mn}
  | m, 1 \rangle = | \cdots \uparrow \stackrel{m}{\uparrow} \downarrow
  \downarrow \cdots\rangle,\ \ \ | m,-1 \rangle = | \cdots \downarrow
  \stackrel{m}{\downarrow} \uparrow \uparrow \cdots \rangle.
\end{equation}
The right-hand sides are expressed in the $S^z$-basis.  Here, $m=0,\pm
1,\pm 2,\ldots$ denotes the soliton position ($m=0$ corresponds to the
center of the spin chain), and $Q=\pm 1$ is the charge of the soliton.
We can define the soliton position, charge, and translation
operators---$\widehat{m}$, $\widehat{Q}$, and $\widehat{T}_n$
respectively---as
\begin{equation} \label{eq-m.op}
  \widehat{m} | m, Q \rangle = m | m, Q \rangle, \ \ \widehat{Q} | m,
Q \rangle = Q | m, Q \rangle, \ \ \widehat{T}_n | m, Q \rangle = | m +
n, Q \rangle,
\end{equation}
where $\widehat{m}$ and $\widehat{Q}$ are Hermitian operators, whereas
$\widehat{T}_{n}$ is unitary with $\widehat{T}_n^\dagger =
\widehat{T}_{-n}$.

The one-soliton approximation is tantamount to considering our system
as containing one domain wall and discarding those terms
in~(\ref{eq-h.primary}) which create additional solitons.  For
example, $H^\perp$ in~(\ref{eq-h.perp}) should be discarded because
these terms will always create solitons, but $H^a$ in~(\ref{eq-h.a})
contains terms which translate the soliton by two sites, which will be
identified with $\widehat{T}_2$ and $\widehat{T}^\dagger_2$.
Projecting~(\ref{eq-h.primary}) onto the one-soliton subspace, we
therefore obtain
\begin{equation} \label{eq-h1sol}
  H_{\text{1-sol}} = \frac{1}{2} J^z + \Delta ( T_2 + T_{-2} ) - b Q
                     m,\ \ \Delta = (J^y - J^x)/4,
\end{equation}
where we have dropped the hats over the operators, and are measuring
energy relative to the fully polarized (ferromagnetic) state.  The
band width $\Delta$ is different from zero only if $J_x\neq J_y$, in
contrast to {\sc afm}s or alternate field configurations (see below).
In what follows, we shall work in a fixed charge sector.  Thus, $Q$ is
effectively a constant and we will put $Q = -1$ to be definite.
Equation~(\ref{eq-h1sol}) is then formally equivalent to a single-band
tight-binding model of an electron in an external electric field.
Solitons play the role of electrons, and a magnetic field the role of
the electric field.  We can now go on to discuss the eigenstates and
energy spectrum, both of which are qualitatively very different in the
finite- and zero-field regimes respectively.  Bloch oscillations can
be derived either semiclassically from the zero-field solution, or
fully quantum-mechanically from the finite-field solutions.  We
discuss each of these in turn.

\subsubsection{Semiclassical Solution}

We will first consider the known physics in zero external field in
order to make contact with previous work on {\sc
afm}s,\cite{villain75} {\sc fm}s,\cite{braun96a} as well as with
semiclassical derivations of {\sc bo}.  For $b=0$, and with periodic
boundary conditions, the eigenstates of~(\ref{eq-h1sol}) describing
the soliton are extended Bloch states labeled by the wave vector $k$:
\begin{equation} \label{eq-kstates}
  | k, Q \rangle = {1 \over \sqrt{ N_{ \text{tot} } } } \sum_m e^{i k
  m} | m,Q \rangle.
\end{equation}
Here, we have set the lattice constant $a$ equal to one.  The periodic
dispersion relation resulting from these eigenstates is given
by\cite{braun96a}
\begin{equation} \label{eq-disp}
  E(k) = \frac{1}{2} J^z + 2 \Delta \cos (2 k).
\end{equation}
This is the ferromagnetic analog of Villain's result\cite{villain75}
for {\sc afm}s.  As mentioned below, Villain's result for the
bandwidth $\Delta$ contains the sum, rather than the difference, of
the transverse couplings.

Semiclassically, we can reproduce the derivation of {\sc bo} as it is
given in conventional electronic treatments.\cite{madelung,mendez93}
In the absence of scattering, the effect of the field-dependent term
$b Q m$ in~(\ref{eq-h1sol}) is to drive the soliton through the band.
The velocity $v(k)$ of this motion is found by
differentiating~(\ref{eq-disp}) with respect to $k$.  On the other
hand, the wave vector $k$ acquires a time dependence through the force
$F = \hbar \dot{k} = b / a$.  Integrating $v(k(t))$ over time then
yields the semiclassical Bloch oscillations.  If $x(t)$ denotes the
soliton position, then
\begin{equation} \label{eq-sem.x}
  x(t)= {\rm const.} - \frac{1}{2} A_B \cos (\omega_Bt),
\end{equation}
with the Bloch amplitude $A_B$ and the Bloch angular frequency
$\omega_B$ given by
\begin{equation} \label{eq-ba.bf}
    A_B = 4 \Delta a / b,\ \ \ \hbar \omega_B = 2 b = 2 g \mu_B B.
\end{equation}

\subsubsection{Quantum Solution (Exact)}

We can compare the above semiclassical derivation of {\sc bo} with a
fully quantum treatment by keeping a nonzero magnetic field in the
Hamiltonian right from the outset.  For a spin-1/2 chain with
$N_{\text{tot}} = 2 N + 1$ sites, and with $b > 0$, we can exactly
diagonalize~(\ref{eq-h1sol}) to yield energy eigenstates
\begin{mathletters}
\label{eq-eigen}
  \begin{equation} | E_m \rangle = \sum_{n=-N}^N C_{m n} | n \rangle,
    \end{equation} \begin{equation} \label{eq-exp} C_{m n} =\langle n|
    E_m \rangle = \frac{1 + (-1)^{m-n}}{2} J_{(m-n)/2}(\alpha),\ \ \
    \alpha = \Delta / b, \end{equation}
\end{mathletters}
where $J_n$ is the ordinary Bessel function of order $n$, and we have
dropped the label $Q = -1$ in the state vectors.  The state $| E_m
\rangle$ is {\em localized} about the lattice site $m$ in the sense
that $\lim_{n \rightarrow \infty} \langle n | E_m \rangle = 0$ for
arbitrary $m$.  We have thus chosen vanishing boundary conditions---a
reasonable choice for localized states.  The degree of localization is
given by the argument of the Bessel function
(see~(\ref{eq-bessel.asympt}) below); strong fields act to localize
the states.  For example, if $b \rightarrow \infty$, then $\alpha
\rightarrow 0$, and only the $m=n$ term contributes to the sum
in~(\ref{eq-eigen}) since $J_n(0) = \delta_{n,0}$.  The eigenstate is
therefore strongly localized.  As the field decreases, the wavepacket
spreads.  The eigenstates at finite $b$ are Wannier-like states and
are thus qualitatively different from the extended Bloch states at
zero field.  This difference is also reflected in the energy spectrum:
\begin{eqnarray}\label{eigenvalue.eq}
   H_{\text{1-sol}} \left| E_m \right\rangle &=& \sum_n C_{m,n} \left[
   \Delta \left( \left| n+2 \right\rangle + \left| n-2 \right\rangle
   \right) + bn \left| n \right\rangle \right] \nonumber\\&=& \sum_n
   \left[ \Delta \left( C_{m,n-2} + C_{m,n+2} \right) + b n C_{m,n}
   \right] \left| n \right\rangle \nonumber\\&=& \sum_n \left[ \Delta
   \frac{2}{\alpha} \left(\frac{m-n}{2}\right) C_{m,n} + b n C_{m,n}
   \right] \left| n \right\rangle = b m \sum_n C_{m,n} \left| n
   \right\rangle = b m \left| E_m \right\rangle,
\end{eqnarray}
and thus 
\begin{equation} \label{eq-ws.ladder}
  E_m = b m.
\end{equation}
The important first term on the third line in~(\ref{eigenvalue.eq})
was obtained using the Bessel function identity\cite{abramowitz72}
$zJ_{\nu - 1}(z) + zJ_{\nu+1}(z) = 2\nu J_{\nu}(z).$ The spectrum
$\{E_m\}$ is {\em discrete}.  It consists of a series of equally
spaced levels with an energy-level spacing given by $b$.  The analog
of this spectrum for electronic systems is known as the Wannier-Stark
ladder.  Hence, for magnetic solitons, the term Wannier-{\em Zeeman}
ladder ({\sc wzl}) seems appropriate and we shall adopt this term
below.

The states $\{ | E_m \rangle \}$ are exact eigenstates only up to
boundary terms:
\begin{mathletters} \label{eq-corr}
\begin{equation} \label{eq-corr.h}
  H_{\text{1-sol}} | E_m \rangle = b m | E_m \rangle + \text{boundary
  terms},
\end{equation}
where a typical boundary term is given by
\begin{equation}  \label{eq-corr.term}
  J_{\frac{m \pm (N + 2)}{2}} (\alpha) \left| \pm N \right\rangle.
\end{equation}
\end{mathletters}
For large $N$ and finite $b$ (hence finite $\alpha$), these boundary
terms are negligible contributions.  The soliton dynamics are not
expected to depend on the boundary conditions for sufficiently large
chains, especially since the soliton eigenstates {$| E_m \rangle$} are
strongly localized by the field.  For $n \equiv | m - N | / 2 \gg 1$,
and fixed, finite $\alpha$, the Bessel function
in~(\ref{eq-corr.term}) can be replaced by its asymptotic
form\cite{abramowitz72}
\begin{equation} \label{eq-bessel.asympt}
  J_n(\alpha) \approx \frac{ 1 }{ \sqrt{ 2 \pi n }} \left( \frac{ e
  \alpha }{ 2 n } \right)^n,\ \ \ (n \gg 1),
\end{equation}
and therefore decays as $n^{-n}$, {\em provided} the argument $\alpha$
remains fixed at some finite value.  If $N \gg 1$, and assuming
$\alpha$ is indeed fixed, then the correction terms in~(\ref{eq-corr})
can be neglected so long as the center of the eigenstate wavepacket is
not near the boundaries of the chain.  Exact numerical diagonalization
of finite chains (see below) indicates that the states $\{ | E_m
\rangle \}$ can be considered eigenstates everywhere except at a
small, $b$-dependent boundary layer on either end of the chain.  The
correction terms can hence be considered surface effects which become
negligible in the thermodynamic limit.

More severe problems arise in the limit of vanishing field ($\alpha
\rightarrow \infty$).  In this limit, the above asymptotic form of the
Bessel functions cannot be used since now {\em both} the order of the
Bessel function ($\propto N$) as well as its argument ($\propto
1/b^z$) diverge.  The appropriate asymptotic form for $J_n(\alpha)$
depends on the ratio $n / \alpha$; for both $\alpha \rightarrow
\infty$ and $n \rightarrow \infty$, we have\cite{abramowitz72}
\begin{equation} \label{eq-besselasym}
  J_n(\alpha) \approx \sqrt{ \frac{1}{2\pi} } \left( \frac{1}{n^2 -
  \alpha^2} \right)^{1/4} \left( \frac{\alpha}{n + \sqrt{n^2 -
  \alpha^2}} \right)^n e^{\sqrt{ n^2 - \alpha^2 }} \ \ \ \ \ (1 \ll
  \alpha < n).
\end{equation}
This expression reduces to~(\ref{eq-bessel.asympt}) in the limit
$\alpha \ll n$.  Equation~(\ref{eq-besselasym}) shows that
$J_n(\alpha)$ is negligible {\em only if} $|n^2 - \alpha^2| \gg 1$.
This effectively implies that no matter how large the system size
becomes, there will always be a field sufficiently small, such that
the present framework fails.  In practice, this becomes problematic
only when discussing the ``Villain limit'' of vanishing field, where,
in any case, soliton collisions must explicitly be considered (see
below).

\begin{figure}
\begin{center}
\epsfclipon \leavevmode \epsfxsize=\textwidth
\epsffile{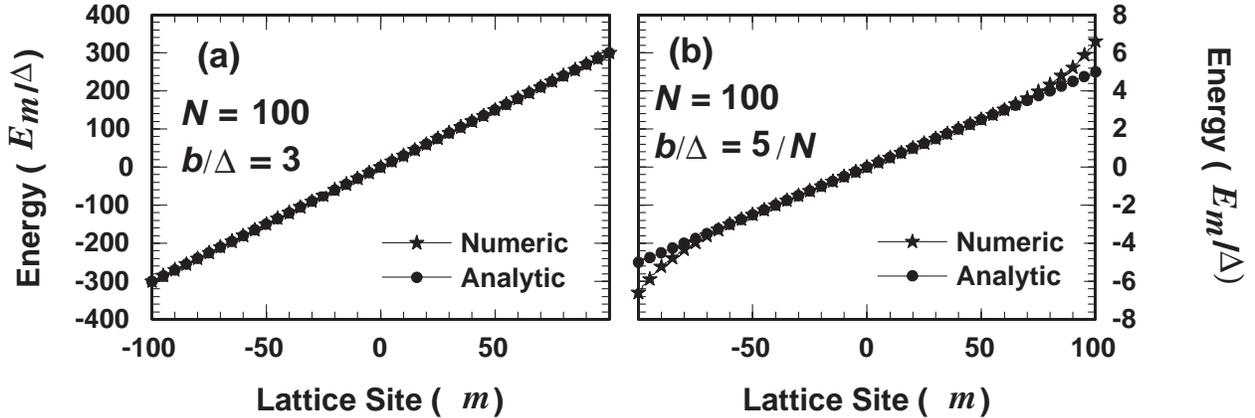}
\end{center}
\caption{(a) Comparison of numerical and analytical results for the
energy spectrum of~(\protect\ref{eq-h1sol}) for moderate field values.
Only every fifth energy level is shown.  (There is one level at each
integer value of $m$.)  For such fields, the analytic spectrum $E_m =
b m$ is essentially exact.  (b) The same plot as in (a), but with a
much smaller field.  Here, about 20\% of the points show significant
deviation between the analytic and numerical results.  As $b$
continues to decrease, this boundary region increases.  For such low
fields, the system exhibits approximate translational invariance and
so a momentum representation becomes more appropriate.}
\label{fig-eig.plot}
\end{figure}

Another approach is to compare the analytic spectrum $E_m = b m$ with
that obtained by numerically diagonalizing finite chains (using
$H_{\text{1-sol}}$).  In Fig.~\ref{fig-eig.plot}(a), we plot the
results for $b /\! \Delta = 3$.  This shows that for such fields, our
analytic expressions are quite good and can be used with confidence.
By contrast, Fig.~\ref{fig-eig.plot}(b) shows a comparison of numeric
and analytic results for $b /\! \Delta = 5/\!N$.  We see here that the
width of the boundary layer has increased greatly.  (The boundary
layer consists of those points which deviate substantially from the
linear analytic result.  In Fig.~\ref{fig-eig.plot}(a), the boundary
layer is not discernible.)  The boundary layer is not a function of
$N$.  It is a function of $b$.  As $N$ increases with $b$ fixed, the
boundary layer therefore becomes less and less important.
Nevertheless, as $b$ tends to zero, the boundary layer increases until
eventually the present framework of localized eigenfunctions must be
abandoned.  This is our first indication of the problems associated
with the $b \rightarrow 0$ limit.  We shall return to this limit below
in connection with the calculation of the dynamical structure factor.
For most of this paper, however, we shall consider either sufficiently
large fields such that Fig.~\ref{fig-eig.plot}(a) is the relevant
scenario, or zero fields, where a dispersive mode with Bloch-like
extended states is the correct description.  These are the two
experimentally relevant regions within the one-soliton approximation.

Assuming a sufficiently large field (in the sense of the previous
paragraph) and neglecting surface effects, the spectrum
of~(\ref{eq-h1sol}) is the {\sc wzl}, $E_m = b m$.  The presence of
the magnetic field thus destroys the continuous band structure
of~(\ref{eq-disp}) and replaces it with an evenly-spaced ladder of
energy levels, with the spacing between adjacent levels given by $b$.
How are {\sc bo} manifested within this fully quantum-mechanical
framework?  To compare with the semiclassical result
of~(\ref{eq-sem.x}) and~(\ref{eq-ba.bf}), we should specify an initial
state and compute the expectation value of $\widehat{m}$ as a function
of time.  Let us keep the initial state arbitrary and write $| \psi(0)
\rangle = \sum_m C_m | E_m \rangle$.  Then,
\begin{eqnarray} \label{eq-qm.mag}
  \langle m(t) \rangle = \langle \psi(t) | m | \psi(t) \rangle =
  \sum_m m | C_m |^2 - {1 \over 2} A_B \, \text{Re} \left( e^{-i
  \omega_B t} \ \sum_m C_m^\ast C_{m-2} \right),
\end{eqnarray}
where Re denotes the real part.  If we assume, for example, that all
$C_m$ are real, and also that $\sum_m C_m C_{m-2} = 1$, then the
oscillating piece of~(\ref{eq-qm.mag}) is identical to the oscillating
piece of~(\ref{eq-sem.x}).  However, if we take as the initial state
one which is completely localized at one lattice site, $| \psi(0)
\rangle = | m \rangle$, then the oscillating piece identically
vanishes.  This situation corresponds to the state vector evolving
symmetrically about both sides of the initial position $m$ in a sort
of breather state.  Thus, the behavior depends sensitively upon the
initial conditions.

Semiclassically, we have seen that {\sc bo} result from the soliton
being pushed unscattered through the band and undergoing Bragg
reflection at the zone boundary.  By contrast, in a full quantum
treatment, {\sc bo} result from the time evolution of a state which is
not a (Wannier-Zeeman) eigenstate of the Hamiltonian.  The frequency
of oscillation is given by twice the energy between adjacent states.
This is a result of the fact that~(\ref{eq-h1sol}), with $b=0$,
contains an intrinsic periodicity of {\em two} lattice constants,
which also accounts for the fact that the Brillouin zone is halved, as
shown by~(\ref{eq-disp}). This effect can be understood
semiclassically in terms of the Berry phase of the spins and is a
result of the spin parity (see Ref.~\onlinecite{braun96} for more
details).

Finally, we note that {\sc bo} are a many-body effect which should be
distinguished from Larmor precession of uncoupled spins in an external
field.  The former yields an oscillating magnetic moment along the
direction of the applied field whereas the latter yields oscillations
in directions transverse to the field.

We have focused on biaxial {\sc fm}s because the materials we have
identified as candidates for observing {\sc bo} are all biaxial {\sc
fm}s.  In the next subsection, we show that {\sc bo} of magnetic
solitons can also exist in uniaxial {\sc fm}s ($J^x = J^y$) if the
field is tilted away from the Ising-axis.  We also show that {\sc bo}
exist in anisotropic {\sc afm}s if an inhomogeneous field is applied.

\subsection{Other Models}

\subsubsection{Ferromagnets} \label{sec-fm}

It is not necessary to have a biaxial {\sc fm} in order to create {\sc
bo}.  In fact, a uniaxial {\sc fm} ($J^x = J^y$) may be preferable
because the band width can then be externally controlled.  For
example, consider~(\ref{eq-h.general}) with ferromagnetic couplings,
with $J^x = J^y$, and with a homogeneous magnetic field along both the
$x$- and $z$-axes ($z$ is still the Ising axis).  The Hamiltonian is
almost the same as~(\ref{eq-h.primary}); the differences are that
$H^a$ now vanishes (because $J^x = J^y$) and there is a new term
coming from the field in the $x$ direction:
\begin{equation}
  H^{a'} = -{1 \over 2} b^x \sum_n (S_n^+ + S_n^-).
\end{equation}
This term gives hopping by one site.  The one-soliton Hamiltonian is
\begin{equation}
  H_{\text{1-sol}}^{\text{uni}} = \frac{1}{2} J^z - \frac{1}{2} b^x (
  T_1 + T_{-1} ) - b^z Q m,
\end{equation}
which is practically identical to~(\ref{eq-h1sol}).  Thus, uniaxial
{\sc fm}s exhibit the {\sc wzl} if the external field is tilted away
from the Ising axis.  The important difference between this and the
biaxial case is that now the strength of {\em both} terms are
adjustable externally.  The energy eigenvalues here are as
in~(\ref{eq-ws.ladder}), but the eigenstates are replaced by
\begin{equation}
  \left| E_m \right\rangle^{\text{uni}} = \sum_n J_{m-n} \left( -b^x /
  b^z \right) | n \rangle.
\end{equation}
Although some of the materials we will identify in the following
sections are reported to have only uniaxial anisotropy, the material
which seems to the best characterized, and for which we provide the
most detailed analysis, is one which is reported to be a {\sc fm} with
biaxial anisotropy.

\subsubsection{Antiferromagnets} \label{sec-afm}

It is more difficult to achieve {\sc bo} in {\sc afm}s because of the
local N\'eel order.  We mentioned above that {\sc bo} can be viewed,
at least semiclassically, as the result of applying a force on a
particle in a band.  How can one apply a force to an antiferromagnetic
domain wall?  The force is given by ${\bf F} = - \nabla \sum_n {\bf
b}_n \cdot {\bf S}_n$. If the external field ${\bf b}$ is homogeneous,
then the force quickly averages to zero over the chain.  However,
applying an {\em inhomogeneous} field produces a net force.
Equation~(\ref{eq-h.primary}) can still be used, but the couplings are
now negative, and $H^I$ must be slightly altered to reflect the
inhomogeneity of the field:
\begin{equation}
  H^I_{\text{\sc afm}} = \left|J^z\right| \sum_n S^z_n S^z_{n+1} -
  \sum b^z_n S^z_n.
\end{equation}
The one-soliton states must also be redefined.  Rather
than~(\ref{eq-mn}), we should write
\begin{equation} \label{eq-mn.afm}
  | m, Q \rangle = | \cdots \uparrow \downarrow \uparrow
  \stackrel{m}{\downarrow} \downarrow \uparrow \downarrow \uparrow
  \cdots \rangle.
\end{equation}
The charge $Q$ can be defined by the first spin at the end of the
chain:\cite{braun96a} $Q=-1$ for spin-up, and $Q=+1$ for spin-down.
Now it is $H^a$ in~(\ref{eq-h.a}) which will always create additional
solitons (and thus should be discarded in the one-soliton
approximation).  Conversely, $H^\perp$ translates solitons.  As a
specific example, we can take a magnetic field along the easy axis
that linearly increases along the chain axis: $b_n^z = n b^z$ (such a
field satisfies Maxwell's equation $\nabla\cdot {\bf B}=0$).
Projecting down to the one-soliton sector, we obtain
\begin{equation} \label{eq-h1sol.afm}
  H_{\text{1-sol}}^{\text{\sc afm}} =\frac{1}{2}|J^z| +
  \Delta^{\text{\sc afm}} ( T_2 + T_{-2} ) - {1 \over 2} b^z Q (-1)^m
  \left(m + {1 \over 2} \right),\ \ \ \Delta^{\text{\sc afm}} =
  \left(J^y + J^x\right) / 4.
\end{equation}
This Hamiltonian is again similar to~(\ref{eq-h1sol}).  The spectrum
again consists of the {\sc wzl}, but due to the antiferromagnetism,
the dependence of the spectrum on the position variable $m$ is
slightly altered:
\begin{equation}
  E_m = {1 \over 2} b^z (-1)^m \left( m + {1 \over 2} \right).
\end{equation}
The eigenstates are also very similar to~(\ref{eq-eigen}).  Only
$\alpha$ must be slightly changed, again to reflect the
antiferromagnetism:
\begin{equation}
  \alpha = (-1)^m \frac{ J^x + J^y }{ 2 b^z },\ \ \ (Q=-1).
\end{equation}
Thus, much of what follows applies also to {\sc afm}s where the
anisotropy can be either uniaxial or biaxial.  One must only replace
the homogeneous field with a linearly increasing field.

In most of what follows, we shall consider biaxial {\sc fm}s with
static and homogeneous fields.  The present subsection, however, shows
that the same analysis can be carried over, almost without change, to
uniaxial {\sc fm}s and to biaxial and uniaxial {\sc afm}s.

\subsection{Conditions for Observation}

Here, we briefly touch on the conditions necessary to observe {\sc bo}
in physical systems. This discussion will have to remain rather vague,
as it requires a detailed knowledge of specific material properties
and, based on that, further theoretical investigations.  Still, we can
list a few essential conditions in general terms, which are very
similar to the ones studied in the context of mesoscopic effects in
electronic systems.\cite{Imry}

First, there should be no Zener transitions (interband tunneling).
The soliton oscillates only when it is reflected from one zone
boundary to the opposite one within the same band.  If the force on
the soliton is too strong, it will gain so much energy at the top of
the band that it will tunnel into a higher energy band. This tunneling
will produce classical linear motion, rather than the quantum
mechanical {\sc bo}.  Such transitions can be neglected if the Bloch
frequency is much larger than the Zener transition rate.  This
effectively puts an upper bound on the field driving the particle. In
the present context, it means that the exchange constant $J^z$ along
the easy axis, a measure of the band gap, should be much larger than
the magnetic field $b$.

Second, there should be no inelastic scattering.  Such events occur,
for example, by emission and absorption of phonons, or via
soliton-soliton interaction.  Inelastic scattering may destroy the
phase coherent motion of the particles necessary for {\sc bo} to
occur.  A detailed investigation into the nature and especially the
magnitude of the spin-lattice coupling in the spin-chains we shall be
discussing below is beyond the scope of the present work, and is, in
any case, probably a matter best determined experimentally (e.g. from
the measured line-width in the structure factor).  It is known,
however, that the spin-lattice couplings are far weaker than the
analogous charge-lattice couplings.  Also, if the soliton density is
low enough, soliton-soliton interactions, being typically of
short-range nature, can be neglected and the results of band theory
are still valid.  In Ising-like spin chains, a low-density requirement
implies that the temperature should be less than the exchange coupling
$J_z$.  Again this can be typically satisfied.

Finally, elastic scattering, such as scattering from static random
impurities, may also be a problem (although typically less
restrictive).  Here, one should consider the Anderson localization
length induced by random disorder in low-dimensional systems.  This
length, which is on the order of the elastic mean free path of the
propagating quasiparticle\cite{Economou} (in the present case, the
soliton), should be greater than the Bloch amplitude, which places a
lower bound on the $B$ field driving the soliton.  However, since the
Bloch amplitude can typically be on the order of the lattice constant,
this poses no severe constraints.

Although the above conditions are demanding, it is quite encouraging
that the presence of extended states of dispersive solitons (i.e. the
Villain mode) has been established experimentally in Ising-like
antiferromagnets.%
\cite{yoshizawa81,boucher85,nagler82,nagler83,adachi81,ajiro89,mogi87}
This suggests that, at least for certain spin-chains, inelastic
scattering and disorder can be neglected to first approximation.  In
the end, the inelastic mean free time $\tau_{\rm in}$ should be
compared with the Bloch frequency $\omega_B$.  Bloch oscillations are
possible if $\omega_B > 1/\tau_{\rm in}$.  Typical values for
$\omega_B$ lie between 40--600\,GHz (see below).

\section{The Dynamical Structure Factor}

In this section, we show that the dynamical structure factor $S^{zz}_b
(q,\omega)$ at finite field contains sharp peaks at integer multiples
of the Bloch frequency $\omega_B$---clear evidence of the {\sc
wzl}. Thus, inelastic neutron scattering, for example, can detect the
{\sc wzl}.  By contrast, we also calculate the dynamical structure
factor $S^{zz}_0(q,\omega)$ at zero field and thus give the
ferromagnetic analog of the Villain mode for {\sc afm}s.

For a translationally invariant system such as the full Hamiltonian
in~(\ref{eq-h.primary}), the dynamical structure factor is defined in
the standard way:\cite{Forster}
\begin{equation}
  S^{zz}(q,\omega) = {1 \over 2 \pi} \int_{-\infty}^\infty \! dt \,
  e^{-i \omega t} \langle \delta S_{-q}^z(0) \delta S_{q}^z(t)
  \rangle,
\end{equation}
where $\delta S_q^z = S_q^z - \langle S_q^z \rangle$.  The Fourier
transform of the spin operator is also defined in the standard way:
for a finite chain with $N_{{\rm tot}} = 2 N + 1$ sites, $S_q^z =
\sum_{n=-N}^N e^{i q n} S_n^z$.  (We continue to set $a=1$.)  If the
eigenbasis $\{ |\psi_m\rangle \}$ is orthonormal and discrete, one may
write
\begin{eqnarray} \label{eq-sqwcalc}
  S^{zz}(q,\omega) &=& \frac{1}{Z} \sum_{m,n} e^{-\beta E_m} \left|
  \langle \psi_n | S_q^z | \psi_m \rangle \right|^2 \delta \biglb(
  \omega - (E_n - E_m) \bigrb) - \left| \left\langle S_q^z
  \right\rangle \right|^2 \delta(\omega),
\end{eqnarray}
where $E_m$ is the energy eigenvalue.

In the following subsection, we give results for $b = 0$ and later
give results for $b \neq 0$.

\subsection{The Dispersive Mode ($b=0$)}

Without a magnetic field, the eigenstates are the Bloch
states~(\ref{eq-kstates}).  Substituting these states, for a fixed
charge of $Q=-1$, into~(\ref{eq-sqwcalc}) yields
\begin{eqnarray}
\label{eq-sqw.bo}
  S_0^{zz} (q,\omega) = \frac{1}{Z} \sum_{k,k'} e^{-\beta E(k)} \left|
  \left\langle k' \right| S_q^z \left| k \right\rangle \right|^2
  \delta \biglb( \omega - E(k') + E(k) \bigrb) - \left|\left\langle
  S_q^z \right\rangle\right|^2 \delta(\omega),
\end{eqnarray}
where $E(k)$ is shown in~(\ref{eq-disp}).  (We neglect the constant
factor of $J^z/2$ since it drops out in the end.)  The partition
function $Z$ can be expressed in terms of the modified Bessel function
of order zero:
\begin{equation}
\label{eq-z.b0}
  Z = \sum_k e^{-\beta E(k)} = N_{\rm tot} I_0 (2 \Delta \beta).
\end{equation}
The matrix elements of $S_q^z$ in the eigenbasis can be found by first
noting that
\begin{equation} \label{eq-s.op}
   S_n^z |m \rangle = \left\{ \begin{array}{rl} -(1/2) | m \rangle,\ &
   m \ge n \\ (1/2) | m \rangle,\ & m < n, \end{array} \right.
\end{equation}
from which it follows that
\begin{equation}
  S_n^z \left| k \right\rangle = \frac{1}{2} \left| k \right\rangle -
  \frac{1}{\sqrt{N_{\rm tot}}} \sum_{m=n}^N e^{i k m} \left| m
  \right\rangle.
\end{equation}

The one-soliton approximation breaks down as $q \rightarrow
0$.\cite{villain75,nagler83,boucher85} If the neutron transfers no
momentum to the system, then the energy will likely go into the
creation of another soliton (actually a soliton-antisoliton pair).
These are not the processes we are interested in here and are not
contained in our approximation.  Rather, we wish to consider the
process where a neutron {\em scatters} the soliton from one $k$-state
to another, which is less and less likely to happen as $q \rightarrow
0$.  Below, we shall give estimates of both the soliton density and
collision rate.

Limiting the discussion to $q \neq 0$, the matrix elements of $S_q^z$
are given by $\langle k' | S_q^z | k \rangle = (e^{iq} \delta_{k',k+q}
- e^{-iqN} \delta_{kk'} ) / (1 - e^{iq})$, and so the modulus squared
becomes
\begin{equation}
  \left| \left\langle k' \right| S_q^z \left| k \right\rangle
    \right|^2 = \frac{1}{4 \sin^2 (q/2)} \left( \delta_{k',k+q} +
    \delta_{kk'} \right).
\end{equation}
From this, we can find the first term in~(\ref{eq-sqw.bo}):
\begin{equation}
\label{eq-3}
  S_0^{zz} (q,\omega) + \left| \left\langle S_q^z \right\rangle
  \right|^2 \delta(\omega) = \frac{1}{4 \sin^2 (q/2)} \left(
  \delta(\omega) + \frac{1}{Z} \sum_k e^{-\beta E(k)} \delta \biglb(
  \omega - E(k+q) + E(k) \bigrb) \right).
\end{equation}
The Dirac delta function can be written as
\begin{equation}\label{eq-delta}
  \delta \biglb( \omega - E(k+q) + E(k) \bigrb) = \delta \biglb(
  \omega + \Omega_q \sin (2k + q) \bigrb),
\end{equation}
\begin{equation}
  \Omega_q = 4 \Delta \sin q,
\end{equation}
where $\Omega_q$ represents an upper bound or cut-off on the
frequency.  Above this frequency, the structure factor vanishes.  This
is a direct result of the one-soliton approximation, which yields a
dispersion relation with a {\em finite} bandwidth.  For a given
momentum transfer $q$, the maximum energy a soliton can absorb (or
emit), while still remaining in the same band, is $\Omega_q$.  The
delta function~(\ref{eq-delta}) can be written in a more usable form
by using the relation $\delta (f(k)) = \sum_n \delta (k - k_n) /
|f'(k_n)|$, where the $k_n$ are the zeroes of $f(k)$.  In our case,
$f'(k) = 2 \Omega_q \cos(2k+q)$ and the $k_n$ are fixed by the
condition
\begin{equation} \label{eq-ast}
  \sin \left(2k_n+q \right) = -\omega / \Omega_q.
\end{equation}
If we substitute these results into~(\ref{eq-3}), and in addition take
the continuum limit ($\sum_k \rightarrow N_{\rm tot} \int \! dk \, / 2
\pi$ and $N_{\rm tot}\rightarrow \infty$), we obtain
\begin{equation} \label{eq-b}
  S_0^{zz} (q,\omega) + \left| \left\langle S_q^z \right\rangle
  \right|^2 \delta(\omega) = \frac{1}{4 \sin^2 (q/2)} \left(
  \delta(\omega) + \frac{N_{\rm tot}}{2 \pi Z} \sum_n \frac{e^{-2
  \beta \Delta \cos (2k_n)}} {\left| 2 \Omega_q \cos (2k_n + q)
  \right|} \right).
\end{equation}
To perform the sum over $n$, we must first solve~(\ref{eq-ast}).
Defining $\omega / \Omega_q \equiv \sin \phi$, where $|\phi| \le
\pi/2$ since $|\omega| \le |\Omega_q|$, (\ref{eq-ast}) is rewritten as
$\sin (2 k_n + q) = - \sin \phi$.  The general solution is
\begin{equation} \label{eq-gen.sol}
  2 k_n + q = (-1)^{n+1} (n \pi + \phi),
\end{equation}
which is valid for arbitrary integer $n$ ($n = 0, \pm 1, \pm 2,
\ldots$).  But not all values of $n$ are allowed; the allowed values
of $n$ must be chosen such that $|k_n| \le \pi$, implying that the
allowed values of $n$ depend on the values of both $\phi$ and $q$,
where $|\phi| \le \pi/2$ and $|q| \le \pi$.  For example, $n=0$ is
always allowed, $n=-3$ is allowed only if $\pi \le (\phi - q) \le
3\pi/2$, and $|n| \ge 4$ is never allowed.  It turns out that for any
$\phi$ and $q$, there are always four values of $n$ which are
allowed---two even values (either $n = 0,2$ or $n = 0,-2$) and two odd
values (either $n = 1,3$, $n = \pm 1$, or $n = -1, -3$).  Actually,
the specific value of $n$ makes no difference; the important point is
whether $n$ is even or odd (and there are always two of each allowed).
For example, from~(\ref{eq-gen.sol}) we can write
\begin{equation}
  \left| 2 \Omega_q \cos (2 k_n + q) \right| = \left|(-1)^n 2 \Omega_q
  \cos \phi \right| = 2 \sqrt{ \Omega_q^2 - \omega^2 }.
\end{equation}
Similarly, we have
\begin{equation}
  -2 \Delta \beta \cos (2 k_n) =-2 \Delta \beta \left[ (-1)^n \cos
  \phi \cos q - \sin \phi \sin q \right] =\frac{\beta}{2} \left[ \pm
  (-1)^{n+1} \sqrt{\Omega_q^2 - \omega^2} \cot q + \omega \right],
\end{equation}
where the top sign ($+$) is for $|q| \le \pi/2$, and the bottom sign
($-$) for $\pi/2 < |q| \leq \pi$.  (These signs will also prove
irrelevant.)  We can use the above two results to write the sum
in~(\ref{eq-b}) as
\begin{equation}
  \sum_n \frac{e^{-2 \beta \Delta \cos (2k_n)}} {\left| 2 \Omega_q
  \cos (2k_n + q) \right|} = \frac{1}{2 \sqrt{\Omega_q^2 - \omega^2}}
  \sum_n \exp \left[ \frac{1}{2} \beta \left( \pm (-1)^{n+1}
  \sqrt{\Omega_q^2 - \omega^2} \cot q + \omega \right) \right].
\end{equation}
We see now that the magnitude of $n$ plays no role; it only matters
whether $n$ is even or odd.  Since there are always two even and two
odd values of $n$, the sum over the exponential becomes a hyperbolic
cosine, and therefore the additional $\pm$ also becomes irrelevant:
\begin{equation}
  \sum_n \exp \left[ \frac{1}{2} \beta \left(\pm (-1)^{n+1}
  \sqrt{\Omega_q^2 - \omega^2} \cot q + \omega \right)\right] = 4
  e^{\beta \omega / 2} \cosh \left( \frac{1}{2} \beta \sqrt{\Omega_q^2
  - \omega^2} \cot q \right).
\end{equation}
Gathering the above results, the structure factor becomes
\begin{equation} \label{eq-sqw0.first}
  S_0^{zz} (q,\omega) + \left| \left\langle S_q^z \right\rangle
  \right|^2 \delta(\omega) = \frac{1}{4 \sin^2 (q/2)} \left(
  \delta(\omega) + \frac{e^{\beta \omega / 2} \cosh \left( \frac{1}{2}
  \beta \sqrt{\Omega_q^2 - \omega^2} \cot q \right) }{ \pi
  I_0(2\Delta\beta) \sqrt{\Omega_q^2 - \omega^2}} \right).
\end{equation}

\begin{figure}
\begin{center}
\epsfclipon \leavevmode \epsfxsize=8.5cm \epsffile{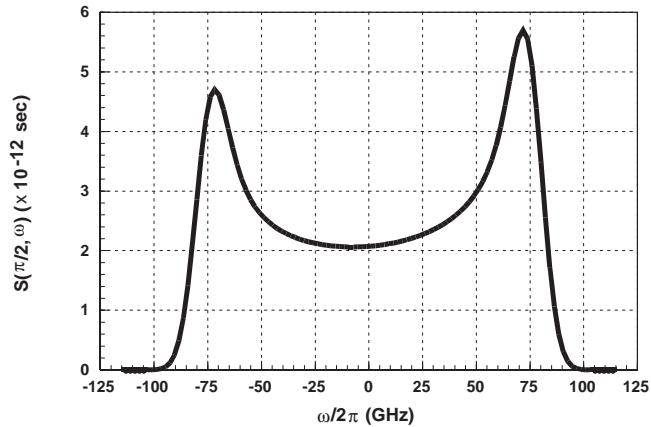}
\end{center}
\caption{A plot of $S^{zz}_0 (q,\omega)$ for $q=\pi/2$, as given
by~(\protect\ref{eq-sqw.b.0}), with $\Delta = 0.925$\,K and $T =
18$\,K.  We have also convoluted $S^{zz}_0 (q,\omega)$ with a
Gaussian, and so the square-root singularities at $\omega = \pm
\Omega_q$ have been rounded, as expected if collisions and
interactions are taken into account.}
\label{fig-villain}
\end{figure}

For the calculation of $| \langle S_q^z \rangle |^2$, we proceed by
noting that, for $q \neq 0$, $\langle k | S_q^z | k \rangle =
-e^{-iqN} / (1 - e^{iq})$.  Since this is independent of $k$, we have
$| \langle S_q^z \rangle |^2 = | \langle k | S_q^z | k \rangle |^2 = 1
/ 4 \sin^2 (q/2)$.  Finally, subtracting this
from~(\ref{eq-sqw0.first}), we obtain $S_0^{zz} (q, \omega)$:
\begin{equation} \label{eq-sqw.b.0}
  S_0^{zz} (q, \omega) = \frac {e^{\beta \omega / 2} \cosh \left(
  \frac{1}{2} \beta \sqrt{\Omega_q^2 - \omega^2} \cot q \right) }{ 4
  \pi \sin^2 (q/2) I_0(2\Delta\beta) \sqrt{\Omega_q^2 - \omega^2} }.
\end{equation}
This expression is plotted in Fig.~\ref{fig-villain} for $\Delta =
0.925$\,K and $T = 18$\,K.  (These parameters are relevant for the
following discussion on the material ${\rm CoCl_2 \! \cdot \!
2H_2O}$.)  The plot shows the frequency dependence at a fixed wave
vector of $\pi/2$.  In addition, we have convoluted the dynamical
structure factor with a Gaussian $R (\omega) = (1 / \sqrt{\pi \sigma})
e^{-\omega^2/\sigma}$ with $\sqrt{\sigma} \approx 4$\,GHz.  Thus, the
square-root singularities at $\omega = \pm \Omega_q$ have been
rounded, which is the expected effect of soliton-soliton collisions
and other interactions we have not taken into account here.  This
structure factor is very similar to that found in {\sc afm}s by
Villain\cite{villain75} and Boucher {\it et al.},\cite{boucher85} who
also worked in the one-soliton subspace, and by Nagler {\it et
al.},\cite{nagler82,nagler83} who worked in the two-soliton subspace.
As argued by Nagler {\it et al.}, similar results should be expected
for all cases where the soliton number stays fixed.

Equation~(\ref{eq-sqw.b.0}) assumes the existence of only one soliton
in the system, whereas there will always be some finite density of
solitons.  Since the (thermal) energy required to create a soliton is
$J^z/2$, the result in~(\ref{eq-sqw.b.0}) may be crudely weighted by a
Boltzmann factor given by\cite{boucher85} $S_0^{zz} (q, \omega)
\rightarrow e^{-\beta J^z \! / 2} S_0^{zz} (q, \omega)$.  A more
proper treatment would be to include soliton interaction, with the
possibility of creation and annihilation of solitons.  Nevertheless,
(\ref{eq-sqw.b.0}) should be qualitatively correct for Ising-like {\sc
fm}s just as the antiferromagnetic analog qualitatively describes the
experimental findings.

An important difference between the result above for {\sc fm}s and the
result for {\sc afm}s is the wave-vector dependence.  The factor of
$\sin^2(q/2)$ in~(\ref{eq-sqw.b.0}) for {\sc fm}s is replaced by
$\cos^2(q/2)$ for {\sc afm}s.  This difference between sine and cosine
is related to the difference in ordering between {\sc fm}s and {\sc
afm}s; for {\sc afm}s, $q=\pi$ is commensurate with the spatial spin
order near the ground state, while for {\sc fm}s, it is $q=0$ which is
commensurate with the ordering.  Thus, one can replace $q$ by $\pi -
q$ in going from {\sc fm}s to {\sc afm}s; this changes $\sin^2(q/2)$
into $\cos^2(q/2)$.  Actually, $\cot q$ and $\Omega_q$ also change
signs, but~(\ref{eq-sqw.b.0}) is invariant under this change.  Also,
the difference in the bandwidth $\Delta$ ($J^y + J^x$ for {\sc afm}s
and $J^y - J^x$ for {\sc fm}s) has been discussed in
Sec.~\protect\ref{sec-afm}.

It should not be so surprising that the result for {\sc fm}s and {\sc
afm}s is so similar in the absence of magnetic fields.  After all, we
have already seen in Sec.~\ref{sec-1.sol.approx} that a variety of
Ising-like models get mapped onto an effective tight-binding model for
solitons in the one-soliton approximation.  For example, in zero
magnetic field, Eq.~(\ref{eq-h1sol}) for {\sc fm}s is formally
identical to Eq.~(\ref{eq-h1sol.afm}) for {\sc afm}s.  Differences
between the two only arise in the presence of a field.  But even then,
one can choose different field configurations for {\sc fm}s and {\sc
afm}s in order to obtain similar structure factors; as shown before,
the {\sc wzl} also exists in {\sc afm}s if an inhomogeneous field is
applied.

Equation~(\ref{eq-sqw.b.0}) shows a divergence as $q \rightarrow 0$.
This should not be viewed as a physical result, but rather as an
indication of the failure of the one-soliton approximation in this
limit.  In contrast, when we consider the $b \neq 0$ case below, we
shall see that the $q \rightarrow 0$ limit is well behaved.  This is
because the soliton states are now localized and the one-soliton
Hamiltonian is no longer translationally invariant on the lattice.
(Thus, $k$ is no longer a good quantum number.)  This localization
should dramatically decrease the collision rate.  In this way, the
magnetic field provides a physical cutoff for the above singularity at
$q=0$.  We will come back to this issue in the following subsection.

For an estimate of the soliton-soliton collision rate, we can employ
the results in Ref.~\onlinecite{boucher85}.  These authors have looked
at the model described by $H_{\text{1-sol}}^{\text{\sc afm}}$
in~(\ref{eq-h1sol.afm}), with $b^z=0$, and so we can use their results
for our ferromagnetic model if we simply substitute $\Delta$ for
$\Delta^{\text{\sc afm}}$.  Following Ref.~\onlinecite{boucher85}, the
soliton density is given by $n_s = e^{-\beta J^z / 2}$ and the soliton
occupation probability by $p(k) = e^{-\beta E(k)} / Z$.  The soliton
velocity, given by the derivative of the dispersion relation, is $v_k
= -4\Delta \sin(2k)$, and the average soliton velocity is defined as
$v_0 = (1 / N_{\rm tot}) \sum_k | v_k | p(k)$, whose evaluation yields
\begin{eqnarray}
   v_0 = \frac{4 \sinh (2 \Delta \beta) } {\pi \beta I_0 (2 \Delta
              \beta)} \stackrel{2 \Delta \beta \ll
              1}{\longrightarrow}& 8 \Delta / \pi.
\end{eqnarray}
The collision rate $\omega_c(k)$ depends on both $n_s$ and $v_k$, and
is given by
\begin{eqnarray}
   \omega_c(k) \approx n_s \sum_{k'} p(k') \left| v_k' - v_k \right|
   \approx n_s v_0 \left[ 1 - \left( 1 - \pi/2 \right) \sin^2(2k)
   \right].
\end{eqnarray}

\subsection{The Wannier-Zeeman Ladder ($b \neq 0$)}

In this subsection, we derive a central result of this paper.  In the
presence of a magnetic field applied along the Ising axis, the
dynamical structure factor will exhibit the signature of the {\sc
wzl}.  We shall also find that the $q \rightarrow 0$ limit is well
behaved, in contrast to the previous subsection on the zero-field
regime.  This, together with the fluctuation-dissipation theorem, will
also enable us to calculate the uniform susceptibility $\chi''
(\omega)$, which provides us with a measure of the magnetization
autocorrelation function.  We begin first with $S^{zz}_b(q,\omega)$,
followed by $\chi'' (\omega)$.

\subsubsection{The Dynamical Structure Factor}

Using the eigenbasis of~(\ref{eq-eigen}) we can write the dynamical
structure factor~(\ref{eq-sqwcalc}) as\cite{foot3}
\begin{eqnarray} \label{eq-sqw-first}
  S^{zz}_b (q,\omega) &=& Z^{-1} \sum_{m,n=0}^{2N} e^{-\beta b m}
  \left| \langle E_{n-N} | S_q^z | E_{m-N} \rangle \right|^2 \delta
  \biglb( \omega - b (n - m) \bigrb) - \left| \left\langle S_q^z
  \right\rangle \right|^2 \delta(\omega).
\end{eqnarray}
Here, we have shifted the origin of our coordinates to one end of the
chain (with a similar shift in the partition function $Z$).  Because
of the simple energy-level structure~(\ref{eq-ws.ladder}), the
partition function is simply given by
\begin{equation} \label{eq-z}
  Z = \frac{ 1 - e^{-\beta b N_{\rm tot}} } { 1 - e^{-\beta b} }
           \stackrel{\beta b N_{\rm tot} \gg 1}{\longrightarrow}
           \frac{1}{1 - e^{-\beta b}}.
\end{equation}
We shall assume $\beta b N_{\rm tot} \gg 1$ in all that follows.

To obtain the matrix elements of $S_q^z$, note first that from
equations~(\ref{eq-s.op}) and (\ref{eq-eigen}), it follows that
\begin{equation} \label{eq-snem}
   S_n^z | E_m \rangle = \frac{1}{2} \left( | E_m \rangle - 2
   \sum_{m'=n}^N C_{mm'} |m'\rangle \right).
\end{equation}
Using this, along with the relations $\langle E_m | m' \rangle =
C_{mm'}$ and $\langle E_m | E_{m'} \rangle = \delta_{mm'}$, we can
find the matrix elements of $S_q^z$:
\begin{eqnarray}
   \langle E_{\bar m} | S_q^z | E_m \rangle &=& \frac{1}{2} N_{\rm
   tot} \delta_{\bar m m} \delta_{q0} -\sum_{n=-N}^N e^{i q n}
   \sum_{m'=n}^N C_{\bar m m'} C_{m m'}.
\label{eq-mat.elem}
\end{eqnarray}
The expansion coefficients $C_{mn}$ are given in~(\ref{eq-exp}).  The
second term on the right side can be brought into a more manageable
form by interchanging the order of the summations:
\begin{equation}
  \sum_{n=-N}^N e^{i q n} \sum_{m'=n}^N C_{\bar m m'} C_{m m'} =
  \sum_{m'=-N}^N C_{\bar m m'} C_{m m'} \sum_{n=-N}^{m'} e^{i q n}.
\end{equation}
Performing the geometric sum over the exponential, the right side is
rewritten as
\begin{eqnarray} \label{eq-piece}
   && \frac{1}{1 - e^{iq}} \sum_{m'=-N}^N \left( e^{-iqN} -
       e^{iq(m'+1)} \right) C_{\bar m m'} C_{m m'} \nonumber \\ &=&
       \frac{1}{1 - e^{iq} } \left( \frac{1+ (-1)^{\bar m - m}}{2}
       \right) \sum_{m' = (m - N)/2}^{(m + N) / 2} \left( e^{-iqN} -
       e^{iq(m + 1 - 2m')} \right) J_{m'}(\alpha) J_{m' + (\bar m -
       m)/2}(\alpha).
\end{eqnarray}
For $N \rightarrow \infty$ (but keeping $m$ finite for now), the sum
over the product of Bessel functions can be performed using the
identity\cite{abramowitz72}
\begin{equation}
  \sum_{k=-\infty}^{\infty} J_k(r) J_{k + n}(\rho) \left\{
  \begin{array}{c} \sin \\ \cos \end{array} \right\} (k\varphi) = J_n
  (R) \left\{\begin{array}{c}\sin\\\cos\end{array}\right\}(n\vartheta)
\end{equation}
where, for all variables real and for $n$ an integer, $R$ and
$\vartheta$ are defined through the relations $R = \sqrt{r^2 + \rho^2
- 2 r \rho \cos \varphi}$, $R \cos \vartheta = \rho - r \cos \varphi$,
and $R \sin \vartheta = r \sin \varphi$.  In particular,
\begin{equation} \label{eq-piece2}
  \sum_{m'=-\infty}^{\infty} e^{-2iqm'} J_{m'}(\alpha) J_{m' + (\bar m
  - m)/2}(\alpha) = \left( -i\, {\rm sign}\, q \right)^{(\bar m -
  m)/2} e^{iq(\bar m - m)/2} J_{(\bar m - m) / 2} (2 \alpha |\sin q|).
\end{equation}
Now we can substitute~(\ref{eq-piece}) and (\ref{eq-piece2})
into~(\ref{eq-mat.elem}), which enables us to write the matrix
elements in closed form as
\begin{eqnarray} \label{eq-mat.elem.fin}
  \left\langle E_{\bar m} | S_q^z | E_m \right\rangle &=&\left(
  \frac{1}{2} N_{\rm tot} \delta_{q0} -\frac{ e^{- i q N} } { 1 - e^{i
  q} } \right) \delta_{\bar m m} \nonumber \\ &-& \frac{ 1 +
  (-1)^{\bar m - m} }{ 2 } \left( -i \, {\rm sign} \, q \right)^{(\bar
  m - m)/2} \frac{ e^{i q (\bar m + m) / 2} }{ 1 - e^{-i q} } J_{(\bar
  m - m)/2} (\zeta),
\end{eqnarray}
where $\zeta = (2 \Delta / b) |\sin q|$.

Taking the modulus squared and shifting the origin as
in~(\ref{eq-sqw-first}), we can evaluate the first term in the
dynamical structure factor:
\begin{eqnarray} \label{eq-sqw.noself}
   S^{zz}_b (q,\omega) + \left|\left\langle S_q^z
   \right\rangle\right|^2 \delta(\omega) &=& \left( \frac{1}{4} N_{\rm
   tot} - 1 - \frac{1}{Z} \sum_{m=0}^{2N} me^{-\beta b m} \right)
   N_{\rm tot} \delta_{q0} \delta(\omega) \nonumber \\ &+& \left(
   \frac{1}{2} - J_0 (\zeta) \frac{1}{Z} \sum_{m=0}^{2N}
   \cos[q(m+1)]e^{-\beta b m} \right) \frac{
   \delta(\omega)}{2\sin^2(q/2)} \nonumber \\ &+&
   \frac{J^2_{\omega/2b}(\zeta) }{4\sin^2(q/2)} \frac{1}{Z}
   \sum_{m,n=0}^{2N} {}^{\!\!\!\!\!\textstyle\prime} \ e^{-\beta b m}
   \delta \biglb( \omega - b(n-m) \bigrb).
\end{eqnarray}
The prime on the final summation indicates that only those terms with
$m - n$ even should be included.  The first of three sums
in~(\ref{eq-sqw.noself}) is proportional to the derivative of the
partition function:
\begin{equation}\label{eq-sum1}
  \frac{1}{Z} \sum_{m=0}^{2N} me^{-\beta b m} = -\frac{1}{b}
  \frac{\partial}{\partial \beta} \ln Z = \frac{1}{e^{\beta b} - 1}.
\end{equation}
The second sum in~(\ref{eq-sqw.noself}) can be written as a geometric
series by writing the cosine as an exponential.  The result is
\begin{eqnarray}\label{eq-sum2}
  \frac{1}{Z} \sum_{m=0}^{2N} \cos[q(m+1)] e^{-\beta b m} &=&
  \frac{(\cos q - e^{-\beta b} ) (1 - e^{-\beta b}) } {\left| e^{-iq}
  - e^{-\beta b} \right|^2},
\end{eqnarray}
For the third and final sum in~(\ref{eq-sqw.noself}), progress can be
made by breaking it up into one containing terms with $n$ and $m$
even, and another containing terms with $n$ and $m$ odd; for the odd
sums, we put $n \rightarrow 2n+1$ and $m \rightarrow 2m+1$, and for
the even sums, $n \rightarrow 2n$ and $m \rightarrow 2m$.  If we then
recombine these two sums, and in addition take $\nu \equiv n-m$, the
final sum in~(\ref{eq-sqw.noself}) can be written as
\begin{equation}\label{eq-int}
  \sum_{m,n=0}^{2N} {}^{\!\!\!\!\!\textstyle\prime} \ e^{-\beta b m}
  \delta \biglb( \omega - b(n-m) \bigrb) \longrightarrow
  \left(1+e^{-\beta b}\right) \sum_{m=0}^N e^{-2\beta b m} \sum_{\nu =
  -m}^{N-m} \delta (\omega - 2 b \nu).
\end{equation}
Now we can interchange the order of the sums by using the identity
\begin{equation}
  \sum_{m=0}^N e^{-2\beta b m} \sum_{\nu = -m}^{N-m} \delta (\omega -
  2 b \nu) = \sum_{\nu = -N}^0 \delta (\omega - 2 b \nu)
  \sum_{m=-\nu}^N e^{-2\beta b m} + \sum_{\nu = 1}^N \delta (\omega -
  2 b \nu) \sum_{m=0}^{N-\nu} e^{-2\beta b m}.
\end{equation}
The geometric sums can now be performed.  Using also the fact that
$\beta b N \gg 1$, (\ref{eq-int}) is written as
\begin{equation}
\sum_{m,n=0}^{2N} {}^{\!\!\!\!\!\textstyle\prime} \ e^{-\beta b m}
  \delta \biglb( \omega - b(n-m) \bigrb) = Z e^{\beta(\omega -
  |\omega|)/2} \sum_{n=-N}^N \delta (\omega - 2 b n).  \label{eq-sum3}
\end{equation}
Substituting~(\ref{eq-sum1}), (\ref{eq-sum2}), and (\ref{eq-sum3})
into~(\ref{eq-sqw.noself}) yields
\begin{eqnarray}\label{eq-sqw2}
   \lefteqn{S^{zz}_b (q,\omega) + \left|\left\langle S_q^z
   \right\rangle\right|^2 \delta(\omega) = \left( \frac{1}{4} N_{\rm
   tot} - \frac{1}{1-e^{-\beta b}} \right) N_{\rm tot} \delta_{q0}
   \delta(\omega)} \nonumber \\&&\mbox{}+ \left( \frac{1}{2} - J_0
   (\zeta) \frac{ \left( \cos q - e^{-\beta b} \right) \left( 1 -
   e^{-\beta b}\right) }{\left| e^{-iq} - e^{-\beta b} \right|^2 }
   \right) \frac{ \delta(\omega)}{2\sin^2(q/2)} +\frac{ e^{\beta
   (\omega - |\omega|) / 2} J^2_{\omega/2b}(\zeta)} { 4 \sin^2 (q/2) }
   \sum_{n=-N}^N \delta(\omega - 2 b n).  \nonumber \\
\end{eqnarray}
The last term is the most interesting one; it will induce transitions
between Wannier-Zeeman levels.  Let us first, however, complete the
calculation by determining $|\langle S_q^z \rangle|^2$. {}
From~(\ref{eq-z}) and (\ref{eq-mat.elem.fin}), we have
\begin{eqnarray} \label{eq-sqwself}
  \left|\left\langle S_q^z \right\rangle\right|^2 &=& \left|
  \frac{1}{2} N_{\rm tot} \delta_{q0} -\frac{e^{-iqN}}{1-e^{iq}}
  \left( 1 - J_0(\zeta) \frac{1 - e^{-\beta b}}{e^{-iq}-e^{-\beta b}}
  \right) \right|^2 =N_{\rm tot} \delta_{q0} \left( \frac{1}{4} N_{\rm
  tot} - \frac{1}{1 - e^{-\beta b}} \right) \nonumber \\ &+& \left( 1
  - 2 J_0(\zeta) \frac{ \left( \cos q - e^{-\beta b} \right) \left( 1
  - e^{-\beta b}\right) }{\left| e^{-iq} - e^{-\beta b} \right|^2} +
  J_0^2(\zeta) \frac{\cosh (\beta b) - 1} { \cosh(\beta b) - \cos q}
  \right) \frac{1}{4\sin^2(q/2)}.
\end{eqnarray}

\begin{figure}
\begin{center}
\epsfclipon \leavevmode \epsfxsize=8.5cm \epsffile{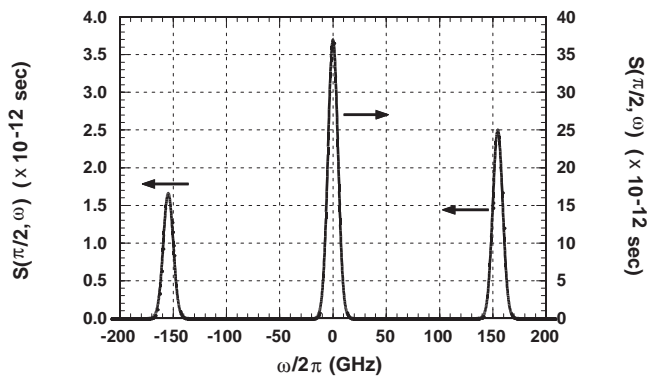}
\end{center}
\caption{A plot of~(\protect\ref{eq-s}) for $q=\pi/2$, with $A_B=a$
and $\beta \omega_B = 0.4$.  For the material ${\rm CoCl_2 \! \cdot \!
2H_2O}$, this corresponds to an applied field of $H^{\text{tot}} =
0.81 \, \text{T}$, $\omega_B / 2\pi = 154 \, \text{GHz}$, and $T = 18
\, \text{K}$.  The peaks at $\omega = \pm \omega_B$ are measured on
the left vertical axis and the peak at $\omega = 0$ is measured on the
right.  The peaks have been broadened by convoluting with a Gaussian
as in~(\protect\ref{eq-sqw.gauss}), with $\protect\sqrt{\sigma}
\approx 40\,{\protect\rm GHz}$.}
\label{fig-sqom}
\end{figure}

\begin{figure}
\begin{center}
\epsfclipon \leavevmode \epsfxsize=8.5cm
\epsffile{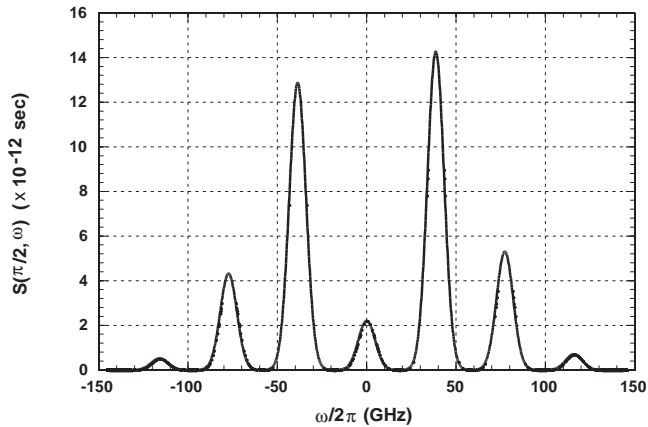}
\end{center}
\caption{The same plot as in Fig.~\protect\ref{fig-sqom}, but with $b$
reduced by a factor of four.  There is a similar decrease in the Bloch
frequency, and a similar {\em increase} in the Bloch amplitude.  The
striking feature is the change in the relative amplitudes of the
peaks, as compared with Fig.~\protect\ref{fig-sqom}.  Here we have
plotted peaks up to three times the Bloch frequency all on the same
scale.}
\label{fig-sqombq}
\end{figure}

Finally, from~(\ref{eq-sqw2}) and (\ref{eq-sqwself}) we obtain the
dynamical structure factor:
\begin{mathletters} \label{eq-s}
\begin{eqnarray} \label{eq-sqw.fin}
  S^{zz}_b(q,\omega) &=& \frac{1}{2} \sum_{n=-N}^N G_n(q)
  \delta(\omega - n \omega_B),
\end{eqnarray} \begin{equation}
  G_0=\frac{ J_0^2 (\zeta) } { \cosh (\beta \omega_B / 2) - \cos q },
\end{equation} \begin{equation}
  G_n = \frac{ J_n^2 (\zeta) }{ 2 \sin^2 (q/2) } \times \left\{
\begin{array}{ccc} 1, && n>0 \\ e^{ n \beta \omega_B }, && n<0,
\end{array} \right.
\end{equation} 
\end{mathletters}
where $\omega_B = 2 b$ is the Bloch frequency.  Equation~(\ref{eq-s})
indicates in particular that inelastic neutron scattering is capable
of {\em mapping the Wannier-Zeeman ladder}.  A soliton initially in
some given Wannier-Zeeman state may be excited to higher states.  The
neutron lineshape intensity for an excitation by $n$ levels, is
essentially given by square of the $n$-th order Bessel function $J_n^2
(\zeta)$.  This means that the {\em relative amplitudes} of the peaks
can be controlled through the argument $\zeta = (2 \Delta / b) |\sin
q|$ by adjusting the external field $b$ (hence also the Bloch
frequency and amplitude).  For example, in Figs.~\ref{fig-sqom} and
\ref{fig-sqombq}, we plot the structure factor as a function of
$\omega$ for $q=\pi/2$.
In these figures, we have convoluted the structure factor with a
Gaussian:
\begin{equation} \label{eq-sqw.gauss}
  S_b^{zz}(q,\omega) \longrightarrow \int^\infty_{-\infty} d\omega'\,
  R(\omega-\omega') S_b^{zz} (q,\omega'),\ \ \ R(\omega) =
  \frac{1}{\sqrt{\pi\sigma}} e^{-\omega^2/\sigma}.
\end{equation}
For the plot in Fig.~\ref{fig-sqom}, we have taken $T=18\,{\rm K}$,
$\Delta = 0.925\,{\rm K}$ (both as in Fig.~\ref{fig-villain}), and $b
= 3.71\,{\rm K}$.  The choice of these numbers is motivated by the
candidate materials to be discussed in the following section.  These
values give a Bloch frequency of about 154\,GHz and a Bloch amplitude
of about one lattice constant.\cite{foot1} For the material ${\rm
CoCl_2 \! \cdot \!  2H_2O}$ discussed in the following section, this
value of $b$ corresponds to an external field of about 0.81\,T, and
the temperature of 18\,K is just above the three dimensional ordering
temperature.  In Fig.~\ref{fig-sqombq}, we use the exact same
parameters as in Fig.~\ref{fig-sqom} {\em except} for $b$, for which
we have put $b = 3.71\,{\rm K}/4 = 0.9275\,{\rm K}$.  The Bloch
frequency is correspondingly reduced by a factor of four, while the
Bloch amplitude is {\em increased} by the same factor.  The striking
feature here is the relative amplitudes of the peaks, as compared with
Fig.~\ref{fig-sqom}; with this smaller field, peaks up to $\omega =
\pm 3 \omega_B$ can be distinguished on the same scale.  The peaks in
both figures away from $\omega=0$ are the signature of the {\sc wzl}.
There exists one peak at every integer multiple of the Bloch
frequency, with an amplitude given by the square of a Bessel function.

We have repeated the above calculation numerically (for finite $N$)
and compared the results with the analytic ones just presented.  For
any given peak at $\omega = n \omega_B$ ($n = 0, \pm1, \pm2, \ldots$),
the numeric results converge to the above analytic ones as the system
size $2N+1$ grows.  If instead we fix $N$, then the numeric results
converge to the analytic results as one moves away from the boundaries
of the chain at $\pm N$.  We may thus conclude that the numeric and
analytic results agree in the thermodynamic limit.

\subsubsection{The Uniform Susceptibility}

It is interesting to note that in contrast to the dispersive mode for
$b=0$, the $q \rightarrow 0$ limit is well behaved when $b\neq 0$:
\begin{equation}
  S_b^{zz} (q \rightarrow 0,\omega) \longrightarrow \frac{
  \delta(\omega) }{ 4 \sinh^2(\beta b / 2) } + \left( \frac{ \Delta }{
  b } \right)^2 \left[ \delta (\omega - \omega_B) + e^{\beta \omega}
  \delta (\omega + \omega_B) \right].
\end{equation}
We have verified this result by performing the calculation at $q=0$
from the outset, as well as by calculating the imaginary part of the
zero wave-vector susceptibility $\chi''(\omega)$ in the Matsubara
formalism, and then using the fluctuation-dissipation theorem:
\begin{equation} \label{eq-susc}
  \chi''(\omega) = \frac{(g \mu_B)^2}{2}(1-e^{-\beta \omega})
  S_b^{zz}(\omega) = \frac{1}{2} \left( \frac{ \Delta }{ B } \right)^2
  \left(1 - e^{-\beta \omega_B} \right) \left[ \delta (\omega -
  \omega_B) - \delta (\omega + \omega_B) \right].
\end{equation}
Here, we have expressed $\chi''(\omega)$ in units of $\mu_B^2 \times
{\rm sec}$.  Equation~(\ref{eq-susc}) represents the response from
only one soliton.  There should be one such factor for each soliton in
the system.  Assume, for example, that one soliton exists every ten
lattice sites for each chain in the sample.  (This requires the Bloch
amplitude to be less than ten lattice sites.)  A single-crystal of
${\rm CoCl_2 \!\cdot\!  2H_2O}$, with a volume of 1\,mm$^3$ will then
contain up to $10^{18}$ solitons.  Thus, the signal can be quite large
and should thus be observable in standard magnetization measurements
(using, for instance, cantilever or {\sc squid} technology).

Again, the structure factor and the susceptibility should be
observable as long as the inelastic scattering rate is less than
$\omega_B$.  Concrete estimates for the soliton collision rate, for
example, in the presence of a $b$-field are even more difficult to
obtain than for $b=0$.  Nevertheless, the rate for sufficiently large
$b$ should be far lower than the rate for $b=0$.  Indeed, when we turn
on the field, the solitons become localized and execute {\sc bo} about
their mean positions. By tuning the field appropriately, the Bloch
amplitude can be kept much smaller than the average distance
(i.e. inverse density) between the localized solitons, and in this
case the soliton-soliton interaction can be expected to be negligible.

We can now go on to compare the zero-field limit of the above
``Wannier-Zeeman'' results with the previous ``dispersive'' results.

\subsection{Discussion of Results ($b \rightarrow 0$)}

As the field decreases, the Bloch amplitude increases.  At some point,
the Bloch amplitude will be equal to the spacing between solitons in
the chain.  At this point, collision effects, which are not included
in our theory, become important and {\sc bo} will be suppressed.  This
means that, for a meaningful comparison between our approximate theory
and experiment, we must consider {\em either} the zero field regime
($B = 0$), {\em or} the regime where $B$ is sufficiently large such
that our one-soliton approximation is justified.

In this sense, the limit of $B \rightarrow 0$ {\em within} the
one-soliton approximation, has no experimental significance. However,
it is still interesting from a technical point of view since this
limit shows some features reminiscent of the classical limit of a
quantum system, in the sense that there might be no {\it pointwise}
convergence.\cite{ballentine90} Other well-known examples are the
harmonic oscillator and a particle in a linear
potential.\cite{ballentine90} In general---and especially when
interference effects are important, as in the {\sc bo} problem---some,
usually {\it ad hoc}, averaging procedure must be employed in order to
obtain a meaningful classical limit.

The problems show up already at the level of the eigenfunctions and
eigenvalues.  At finite field, the eigenfunctions are the localized
Wannier states~(\ref{eq-exp}), $\langle n | E_m \rangle = [1 +
(-1)^{m-n})] J_{(m-n)/2}(\alpha)/2$, which satisfy vanishing boundary
conditions.  But at $b=0$, the extended Bloch states $\langle n | k
\rangle = e^{ikn} / \sqrt{N_{\rm tot}}$ satisfy periodic boundary
conditions.  It is thus not too surprising that as $b \rightarrow 0$,
the Wannier states do not converge pointwise to the Bloch states.  The
same holds for the energy eigenvalues; the band structure cannot be
recovered from the {\sc wzl} in the $b \rightarrow 0$ limit.

Similar remarks apply to the partition functions.  Keeping a finite
system size $N_{\rm tot} = 2 N + 1$, the partition functions for
finite and zero fields are respectively given by
\begin{equation}
  Z_b = \sum_{m=-N}^N e^{-\beta b m} = \frac{ \sinh( \beta b N_{\rm
  tot} / 2 ) }{ \sinh( \beta b / 2 ) },
\end{equation} \begin{equation}
  Z_0 = \sum_{k=-\pi}^\pi e^{-2 \Delta \beta \cos( 2 k ) } = N_{\rm
  tot} I_0(2 \Delta \beta)\, .
\end{equation}
These expressions imply that the $b \rightarrow 0$ limit must be
coupled to the $N_{\rm tot} \rightarrow \infty$ limit.  For example,
if we are in the regime where $\Delta \beta \ll 1$ and $\beta b \ll
\beta b N_{\rm tot} \ll 1$, then to second order in both $\Delta\beta$
and $\beta b$, $Z_b$ converges to $Z_0$ provided we make the
identification $b N_{\rm tot} = 2 \sqrt{6} \Delta$.  On the other
hand, if we are in the regime $\Delta \beta \ll 1$ and $\beta b \ll 1
\ll \beta b N_{\rm tot}$, the partition functions cannot be matched.

\begin{figure}
\begin{center}
\epsfclipon \leavevmode \epsfxsize=8.5cm \epsffile{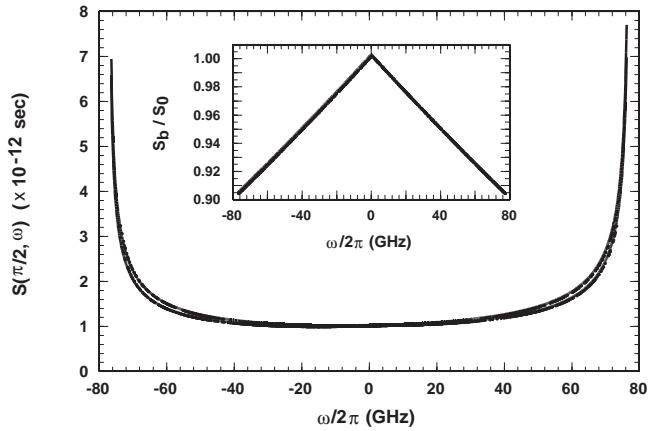}
\end{center}
\caption{A plot of both $\lim_{b \rightarrow 0} S_b^{zz}
(\pi/2,\omega)$ in~(\protect\ref{eq-s.lim}), and $S_0^{zz}
(\pi/2,\omega)$ in~(\protect\ref{eq-sqw.b.0}).  Good agreement is
obtained, but there is some discrepancy as shown in the inset, which
shows the ratio of the two functions~(\protect\ref{eq-ratio}).  This
should be a constant (equal to one) if the two results are the same.
The source of this discrepancy is discussed in the text.}
\label{fig-sqw.comp}
\end{figure}

Next, we look at the dynamical structure factor~(\ref{eq-s}) which we
rewrite here in a slightly different form:
\begin{eqnarray} \label{eq-sqw.fin2}
  S^{zz}_b(q,\omega) =\frac{ J^2_{\omega/2b}(\zeta)
  e^{\beta(\omega-|\omega|)/2}} { 4 \sin^2(q/2) } \sum_{n=-N}^N
  \delta(\omega - 2bn) -\frac{ J^2_0 (\zeta) }{ 4 \sin^2(q/2) } \frac{
  \cosh (\beta b) - 1 }{ \cosh (\beta b) - \cos q } \delta(\omega).
\end{eqnarray}
For $b \rightarrow 0$, the second term tends to zero and so we need
only concern ourselves with the first term.  The Bessel function $
J^2_{|\omega|/2b} \left( \zeta \right) = J^2_{|\omega|/2b} \left(
|\Omega_q| / 2 b \right)$ can be expanded in its asymptotic form for
$b \rightarrow 0$.  If $|\omega| < |\Omega_q|$, then
\begin{equation} \label{eq-sin.rapid}
  J^2_{|\omega|/2b} \left( \frac{|\Omega_q|}{2b} \right) \approx
  \frac{4b}{\pi\sqrt{\Omega_q^2 - |\omega|^2}} \sin^2 \left[
  \frac{1}{2b} \left( \sqrt{\Omega_q^2 - |\omega|^2} - |\omega|
  \arccos \frac{ |\omega| }{ \Omega_q } \right) + \frac{\pi}{4}
  \right].
\end{equation}
It is the rapidly oscillating factor which prevents the pointwise
convergence in the $b \rightarrow 0$ limit; some averaging procedure
is required.  To illustrate this, let us crudely replace the
oscillatory factor with 1/2 (which is accurate only for $\omega^2 \ll
\Omega_q^2$):
\begin{equation}\label{eq-sin.rapid.avg}
  \overline{J^2_{|\omega|/2b} \left( \frac{|\Omega_q|}{2b} \right)}
  \approx \frac{2b}{\pi\sqrt{\Omega_q^2 - |\omega|^2}}.
\end{equation}
In the $b \rightarrow 0$ limit, the sum over $n$
in~(\ref{eq-sqw.fin2}) can be made continuous:
\begin{equation} \label{eq-delta.cont}
  \sum_{n = -N}^N \delta (\omega - 2bn) =\frac{1}{2} \sum_{n =
  -bN}^{bN} \delta \left( \frac{1}{2}\omega - n \right)
  \longrightarrow \frac{1}{2b} \int_{-bN}^{bN} dn \, \delta \left(
  \frac{1}{2}\omega - n \right) = \left\{ \begin{array}{cl} 1/2b,\ &
  |\omega| < 2bN \\ 0,\ & |\omega| > 2bN.  \end{array} \right.
\end{equation}
Inserting~(\ref{eq-sin.rapid.avg}) and (\ref{eq-delta.cont}) into
(\ref{eq-sqw.fin2}), and assuming $|\omega| < |\Omega_q|$, we can
write
\begin{eqnarray} \label{eq-s.lim}
 \lim_{b \rightarrow 0} \overline {S_b^{zz}(q,\omega)} \approx
  \frac{e^{\beta (\omega - |\omega|) / 2} }{ 4 \pi \sin^2(q/2)
  \sqrt{\Omega^2_q - \omega^2} } \neq S_0^{zz}(q,\omega).
\end{eqnarray}
Although the last two expressions are not equal, they are nevertheless
quite similar.  In Fig.~\ref{fig-sqw.comp}, we plot the two results,
Eqs.~(\ref{eq-s.lim}) and (\ref{eq-sqw.b.0}), using the same
parameters as in the previous figures.  The small deviations can be
traced back to the rapidly oscillating sine squared function
in~(\ref{eq-sin.rapid}).  We have replaced this by the constant factor
of 1/2.  This is not quite proper since the period of the oscillation
is a function of $\omega$ (and not just a constant).  Near the cut-off
frequency $\Omega_q$, this error becomes the most apparent because the
rest of the function in~(\ref{eq-sin.rapid}) is also a rapidly varying
function (it is tending to a square root singularity).  In the inset
of Fig.~\ref{fig-sqw.comp}, we have plotted the ratio
\begin{equation} \label{eq-ratio}
  \frac{\lim_{b \rightarrow 0}\overline{ S_b^{zz} (\pi/2,\omega)}}
  {S_0^{zz} (\pi/2,\omega)} \approx \frac{e^{-\beta|\omega|/2} I_0(2
  \Delta \beta)} {\cosh \left( \frac{1}{2}\beta \sqrt{\Omega_q^2 -
  \omega^2} \cot q\right)},
\end{equation}
as a function of $\omega$ for the same values of $\Delta$, $\beta$,
and $q$ used in previous plots.  Near $\omega = 0$ the agreement is
best (the ratio is near unity). But as $\omega \rightarrow \Omega_q$,
the agreement becomes progressively worse.  Nevertheless, this rather
crude treatment achieves reasonable agreement---only about 10\% error
at its worst.  The agreement improves if we choose $q \neq \pi/2$, but
worsens as the product $\Delta \beta$ grows.  (But $\Delta \beta \ll
1$ is the regime of interest.)

This marks the end of the theoretical development.  In the following
section, we shall concentrate on various materials we believe are good
candidates for observing the {\sc wzl}.  Specifically, we shall see
that {\sc bo} can exist in certain ferromagnetic Ising-like salts,
with frequencies on the order of 150\,GHz.

\section{Candidate Materials}
\label{sec-exp}

We have identified four candidate materials for observing {\sc bo} and
the {\sc wzl} in purely magnetic systems.  The materials are all
Ising-like {\sc fm}s, and consist of chains of magnetic ions, with
effective spin-1/2, separated by spacer material.  We focus mainly on
${\rm CoCl_2 \!  \cdot \!  2H_2O}$ (Ref.~\onlinecite{torrance69}), but
give also a brief discussion on the potentially more promising, but
less well characterized, ${\rm CoCl_2 \! \cdot \! 2NC_5H_5}$
(Ref.~\onlinecite{takeda71}), ${\rm [(CH_3)_3 NH] CoCl_3 \! \cdot \!
2H_2O}$ (Ref.~\onlinecite{takeda81}), and ${\rm [(CH_3)_3 NH] FeCl_3
\! \cdot \!  2H_2O}$ (Ref.~\onlinecite{samp4}).

\subsection{${\rm CoCl_2 \! \cdot \!  2H_2O}$}

In ${\rm CoCl_2 \! \cdot \! 2H_2O}$, the magnetic Co ions form chains
along the $c$-axis.  The coupling is ferromagnetic between ions in the
same chain (we consider interchain exchange below).  The exchange
anisotropy is such that the $b$-axis is an easy-axis.  The work of
Ref.~\onlinecite{torrance69} confirms unambiguously that the
Ising-like spin-1/2 Hamiltonian of~(\ref{eq-h.primary}) describes this
system very well.  In Table~\ref{tab-tink}, we list the material
parameters of this ferromagnetic salt.  (We have taken the crystal
$b$-axis to coincide with the $z$-axis of the Cartesian coordinate
frame.)  We also list two antiferromagnetic interchain couplings.  (We
follow Ref.~\onlinecite{torrance69} in neglecting the small non-Ising
part of the interchain exchange.)  We shall consider this interchain
coupling in a mean field treatment by considering the total field at a
given site to be the sum of the externally applied field and some
internal field due to interchain exchange.

\begin{table}
\begin{tabular}{ccccc}
  Parameter & \multicolumn{4}{c}{Value} \\ & CCH\tablenote{${\rm
  CoCl_2 \!\cdot\! 2H_2O}$ (Ref.~\protect\onlinecite{torrance69}).}  &
  CCN\tablenote{${\rm CoCl_2 \!\cdot\! 2NC_5H_5}$
  (Ref.~\protect\onlinecite{takeda71}).}  & CoTAC\tablenote{${\rm
  [(CH_3)_3NH]CoCl_3 \!\cdot\! 2H_2O}$
  (Ref.~\protect\onlinecite{takeda81}).}  & FeTAC\tablenote{${\rm
  [(CH_3)_3NH]FeCl_3 \!\cdot\! 2H_2O}$
  (Ref.~\protect\onlinecite{samp4}).}  \\ \hline $J^z$ & 18.3\,K &
  10\,K & 14.2\,K & 17.4\,K \\ $J^y - J^x$ & 3.7\,K & --- & --- & ---
  \\ $J^y + J^x$ & 5.6\,K & --- & --- & --- \\ $J_1^z$ & $-4.6$\,K &
  $-3.4$\,K & 0.18\,K & $-0.02$\,K \\ $J_2^z$ & $-0.9$\,K & ---
  &$-10^{-3}$\,K & 0.00\,K \\ $a$ & 3.55\,\AA & 3.66\,\AA & 3.63\,\AA
  & 3.68\,\AA \\ $g$ & 6.81 & 5.49 & 6.54 & 7.49 \\ $T_{\rm 3D}$ &
  17.2\,K & 3.17\,K & 4.14\,K & 3.12\,K
\end{tabular} 
\caption{Relevant parameters for the candidate materials discussed in
the text.  A dashed entry (---) means that no value was given in the
references.}
\label{tab-tink}
\end{table}

Let us first neglect the interchain interaction and consider just a
single chain in the presence of a static and homogeneous field along
the $z$-axis.  Then, the one-soliton approximation should be valid if
the external field $b^z = g \mu_B H_{\rm ext}^z$ is less than the
Ising exchange coupling $J^z$.  For the parameters given in
Table~\ref{tab-tink}, this puts a restriction on the field strength of
$H_{\rm ext}^z < 4\, {\rm T}$.  For example, if we apply a 0.81 Tesla
field---comfortably below this upper bound---then the Bloch amplitude
and frequency are given from~(\ref{eq-ba.bf}) as
\begin{equation}
\label{eq-abcocl}
  A_B = a \frac{ J_y - J_x }{ g \mu_B H_{\rm ext}^z } \approx a, \ \ \
  \ \ \ \ \frac{\omega_B}{2 \pi} = \frac{ g \mu_B H_{\rm ext}^z } {
  \hbar \pi } \approx 154 \,{\rm GHz}.
\end{equation}
This amplitude is small enough to impede the destructive influence of
any scattering events, and the frequency falls within the capabilities
of neutron scattering.  This is therefore an encouraging result.

\begin{figure}
\begin{center}
\epsfclipon \leavevmode \epsfxsize=8.5cm \epsffile{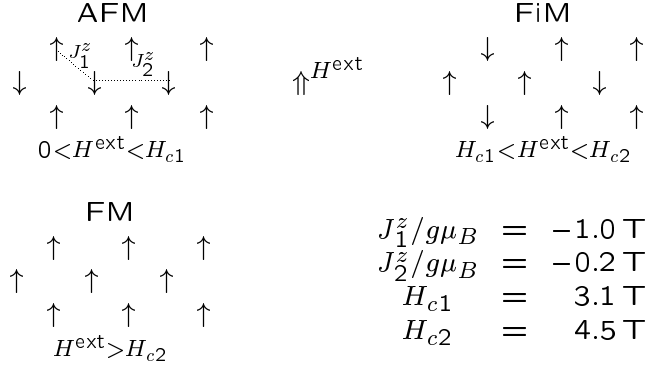}
\end{center}
\caption{The three phases of ${\rm CoCl_2 \! \cdot \! 2H_2O}$ for $T <
17.2$\,K.  The spin chains run perpendicular to the page and $J_1^z$
and $J_2^z$ are the interchain antiferromagnetic couplings.  These
couplings can be considered internal fields, and so they too affect
the Bloch oscillations.  In fact, Bloch oscillations can potentially
exist in all three phases.  This figure has been adapted from
Ref.~\protect\onlinecite{torrance69}.}
\label{fig-phases}
\end{figure}

\begin{figure}
\begin{center}
\epsfclipon \leavevmode \epsfxsize=8.5cm \epsffile{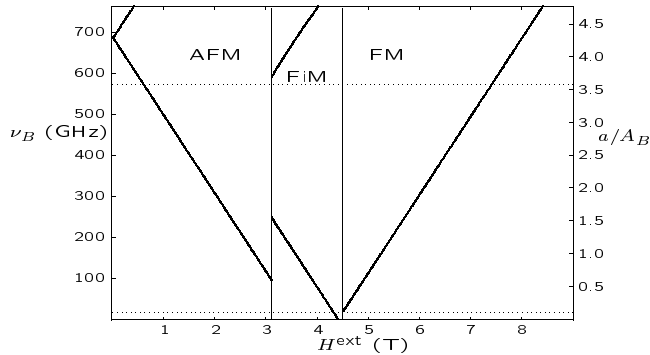}
\end{center}
\caption{A plot of Bloch frequency and inverse Bloch amplitude in
${\rm CoCl_2 \! \cdot \! 2H_2O}$ as a function of external field below
the three dimensional ordering temperature of about 17\,K.  The
discontinuous jumps in the curves are a result of transitions from
anti-, to ferri-, and finally to ferromagnetic order of the chains
relative to each other.  Ferromagnetic order is always maintained {\em
within} each chain.  The horizontal dotted lines denote upper and
lower bounds, beyond which the present analysis should not be expected
to hold.}
\label{fig-bloch}
\end{figure}

Due to the antiferromagnetic interchain couplings $J_1^z$ and $J_2^z$,
the material undergoes a three-dimensional ordering transition at
about 17\,K.  Below this temperature, and in zero external field,
there still exists ferromagnetic order within each chain, but the
chains are ordered antiferromagnetically with respect to each other.
As the field is turned on there are successive transitions from
antiferromagnetic, to ferrimagnetic, and finally to ferromagnetic
order at fields $H_{c1}$ and $H_{c2}$ respectively.  At all times, the
intrachain order is ferromagnetic.  This is depicted in
Fig.~\ref{fig-phases}, where we also list the interchain coupling
values\cite{torrance69} and the critical
fields.\cite{narath66,torrance69}
When determining the Bloch amplitude and frequency, one should also
include these internal fields.  For example, a Bloch amplitude of one
lattice constant, which results from a {\it total} field of $H_{\rm
tot} = 0.81$\,T, can be realized in all three phases when these
internal fields are taken into account.  Explicitly, we have (for
$H_{\text{tot}} = 0.81$\,T)
\begin{mathletters} \begin{eqnarray}
  {\sc \text{afm}}:\ \ \ H_{\downarrow}^{\rm tot} = -H^{\rm ext} -
  4J_1^z/g\mu_B + 2J_2^z/g\mu_B\ \ \ \Longrightarrow H^{\rm ext} = 2.8
  \ {\rm T},
\end{eqnarray} \begin{eqnarray}
  {\sc \text{fim}}:\ \ \ H_{\downarrow}^{\rm tot} = -H^{\rm ext} -
  4J_1^z/g\mu_B - 2J_2^z/g\mu_B\ \ \ \Longrightarrow H^{\rm ext} = 3.6
  \ {\rm T},
\end{eqnarray} \begin{eqnarray}
  {\sc \text{fm}}:\ \ \ H_{\uparrow}^{\rm tot} = H^{\rm ext} +
  4J_1^z/g\mu_B + 2J_2^z/g\mu_B\ \ \ \Longrightarrow H^{\rm ext} = 5.2
  \ {\rm T}.
\end{eqnarray} \end{mathletters}
The notation $H_\downarrow^{\rm tot}$, for example, denotes the total
field at a chain with spin down, where ``down'' is defined as being
opposite to the external field.  Thus, in the ferromagnetic phase, all
chains are spin up.  In Fig.~\ref{fig-bloch} we plot the resulting
predictions for the Bloch frequency and (inverse) Bloch amplitude as a
function of external field in all three phases.
The antiferromagnetic and ferrimagnetic phases show two curves because
these phases have both spin up and spin down chains, and these chains
each feel a different field.  The dotted horizontal lines are bounds,
outside of which the results become equivocal; near the upper bound,
the total field becomes comparable to the Ising exchange constant
$J^z$; near the lower bound, the amplitude becomes too large, so that
scattering effects should probably be taken into account.  There is
however, a fairly large intermediate range of over 400\,GHz where the
effect should be noticeable.  We hope the present work can motivate
some experimental investigation into this material which seems to have
disappeared from the attention of contemporary research.

\subsection{Other Materials}

Another material we wish to mention here is ${\rm CoCl_2 \! \cdot \!
2NC_5H_5}$ (Ref.~\onlinecite{takeda71}).  This material differs from
the one above only by the spacer material---pyridine molecules rather
than water.  Pyridine is a larger molecule than water and so the
magnetic chains are further apart by about a factor of 1.7 (9.4\,\AA\
versus 5.5\,\AA\ for the water spacer).  This material is thus a
better one-dimensional material than ${\rm CoCl_2 \! \cdot \! 2H_2O}$.
The three dimensional ordering temperature is about 5.4 times smaller
than it is in ${\rm CoCl_2 \!  \cdot \! 2H_2O}$ (3.17\,K rather than
17\,K).  The material parameters are summarized in
Table~\ref{tab-tink}.  Because the experimental work on this material
seems less extensive than that on ${\rm CoCl_2 \! \cdot \!  2H_2O}$,
we have not made any predictions for {\sc bo} in this material.  But
due to the reduced three-dimensional ordering temperature, this
material should be a better candidate for observing {\sc bo} and the
{\sc wzl}.  Even if there really is no transverse anisotropy ($J_y =
J_x$), {\sc bo} could still be induced by simply applying a transverse
field in addition to the one along the Ising-axis (see
Sec.~\ref{sec-fm}).

Most of the preceding paragraph applies even more emphatically to the
final two materials listed in Table~\ref{tab-tink}, particularly to
FeTAC.\cite{samp4} The small magnitude of the interchain couplings and
the lower three-dimensional ordering temperature indicate that these
materials may be quite suitable for {\sc bo}.  We have again chosen
not to provide predictions for this compound since we believe further
material characterization is necessary.

In summary, we have shown in this section that there are a number of
materials which may exhibit a dispersive soliton mode as well as {\sc
bo}.  We have not discussed any {\sc afm} chains because we have been
unable to identify any with the appropriate material parameters such
that the Bloch frequency and amplitude fall within experimentally
accessible regimes.  Should any such chains exist, Sec.~\ref{sec-afm}
shows that {\sc bo} may exist under an applied inhomogeneous magnetic
field.

\section{Summary and Outlook}

In this work, we have shown that {\sc bo} of magnetic solitons occur
in anisotropic spin-1/2 chains.  Although we have focused primarily on
biaxial Ising-like {\sc fm}s, we have shown that {\sc bo} can also
occur in uniaxial {\sc fm}s by applying a transverse field in addition
to the longitudinal field. We have also shown that {\sc bo} may exist
in Ising-like {\sc afm}s by applying an inhomogeneous field.

We have been mainly envisioning a neutron scattering experiment in
this work because the dynamical structure factor shows clear evidence
of the {\sc wzl}; it contains sharp peaks at integer multiples of the
Bloch frequency.  At zero wave vector, all the peaks vanish except for
the one at zero frequency and those at the Bloch frequency $\pm
\omega_B$.  Thus, a measurement of the magnetization autocorrelation
function (magnetic susceptibility) about the Bloch frequency should
also detect the {\sc wzl}.

Several materials are promising candidates for observing {\sc bo} and
the {\sc wzl}.  We have chosen to focus our estimates on the
one-dimensional salt ${\rm CoCl_2 \! \cdot \! 2H_2O}$.  Although this
material is not an ideal one-dimensional substance (better ones have
been identified above), {\sc bo} of amplitude one lattice constant
(about 3.6\,\AA) and frequency of about 154\,GHz are possible with
applied fields of a few Tesla.  The other materials we have mentioned
are less well characterized than the one just described.  However, the
much smaller interchain coupling indicates that they are better
one-dimensional samples---a statement further supported by their
three-dimensional ordering temperature, which is much lower than ${\rm
CoCl_2 \! \cdot \! 2H_2O}$.  It is possible that these materials
contain only uniaxial anisotropy.  If so, {\sc bo} can be achieved by
tilting the external field away from the Ising axis.

A question we are currently investigating is what effect
soliton-soliton interactions have on the {\sc wzl}, as well as the
related question on the influence of the higher-soliton states.  The
{\sc wzl} and {\sc bo} should survive if the number of solitons is
conserved.  In Ref.~\onlinecite{nagler83}, a study is presented of
soliton dynamics in the two-soliton sector (but in zero applied
field).  If one enforces periodic boundary conditions, then only even
numbers of solitons can exist.  But quantities calculated in the
thermodynamic limit should be independent of the boundary conditions
employed.  Therefore, as expected, this work found practically the
same result for the dynamical structure factor as Villain did working
in the one-soliton sector.  However, the two-soliton sector brings
with it an opportunity to directly detect the {\em coherent
oscillation of solitons}.  For example, two solitons can form a bound
state which should be identified as a magnon in spin-1/2 chains.
Multiple magnon bound states can then be formed in which a cluster of
adjacent spins are all flipped relative to the majority of the
ferromagnetically aligned spins in the chain.  Indeed, these are
precisely the excitations measured in the work of
Ref.~\onlinecite{torrance69} which concerns the optical excitation of
multiple-magnon bound states.  An enticing scenario exists if the ends
of these clusters also undergo Bloch oscillation.  Rather than having
these excitations thermally created, as we have been assuming above,
one can then optically create these excitations {\it coherently} by
infrared radiation.  The resulting Bloch oscillations will then also
be coherent, and this may be detected, for example, by looking for
coherent emission of magnetic dipole radiation in the microwave
regime.  In this scenario, the magnetic Bloch oscillator is an emitter
of coherent microwave radiation.  This is essentially the analog of
the electronic {\sc bo} experiments, were the charge carriers are
optically excited with visible light, and the electron dipole
oscillations radiate in the submillimeter regime.  There are no
exciton effects in our spin chains and so the detection of the
magnetic radiation would be a clear signal of magnetic Bloch
oscillations.  This intriguing problem will be the subject of a future
publication.

\section*{Acknowledgements}

We thank Guido Burkard, Alain Chiolero, Jack Harris, Bruce Normand,
and Eugene Sukhorukov for discussion and helpful comments.  We also
acknowledge the hospitality of the ITP, Santa Barbara, where the
initial stage of this work was completed.  This work was funded by the
Swiss NSF, the US NSF under grant number PHY94-07194, and the Canadian
NSERC.


\begin{thebibliography}{10}
\bibitem{foot1} An oscillation amplitude of less than one lattice
constant may present conceptual difficulties in the semiclassical
description of a particle following a definite trajectory in
spacetime.  In a fully quantum treatment, such difficulties do not
arise because the description is not in terms of definite
trajectories, but rather in terms of a quantum expectation value for
the soliton's position (which can assume values other than integer
multiples of the lattice constant).
\end{thebibliography}
\end{document}